\documentclass[aps,prl,twocolumn,amsfonts,nobibnotes,superscriptaddress,showpacs]{revtex4-1}

\usepackage{dcolumn}
\usepackage{amsmath}
\usepackage{amssymb}
\usepackage{graphicx}
\usepackage{bm}
\usepackage{color}

\begin{document}

\title{Lattice-mediated magnetic order melting in TbMnO$_3$}

\vspace{2cm}

\author{E. Baldini$^{*}$}
	\affiliation{Institute of Physics and Lausanne Center for Ultrafast Science (LACUS), \'Ecole Polytechnique F\'{e}d\'{e}rale de Lausanne, CH-1015 Lausanne, Switzerland}
	\affiliation{Institute of Chemical Sciences and Engineering and Lausanne Center for Ultrafast Science (LACUS), \'Ecole Polytechnique F\'{e}d\'{e}rale de Lausanne, CH-1015 Lausanne, Switzerland}
	\affiliation{Department of Physics, Massachusetts Institute of Technology, Cambridge, Massachusetts, 02139, USA}
		
\author{T. Kubacka}
	\affiliation{Institute for Quantum Electronics, Eidgen\"ossische Technische Hochschule (ETH) Z\"urich, CH-8093 Z\"urich, Switzerland}	
	
\author{B. P. P. Mallett}
	\affiliation{Department of Physics, University of Fribourg, Chemin du Mus\'ee 3, CH-1700 Fribourg, Switzerland}	
	
\author{C. Ma}
	\affiliation{Department  of  Chemical  Physics, University of Science and Technology of China, Hefei 230026, China}	
	
\author{S. M. Koohpayeh}
	\affiliation{Institute for Quantum Matter, Department of Physics and Astronomy, Johns Hopkins University, Baltimore, Maryland 21218, USA}	
	
\author{Y. Zhu}
	\affiliation{Department of Condensed Matter Physics, Brookhaven National Laboratory, Upton, New York 11973, USA}	
	
\author{C. Bernhard}
	\affiliation{Department of Physics, University of Fribourg, Chemin du Mus\'ee 3, CH-1700 Fribourg, Switzerland}	
	
\author{S. L. Johnson}
	\affiliation{Institute for Quantum Electronics, Eidgen\"ossische Technische Hochschule (ETH) Z\"urich, CH-8093 Z\"urich, Switzerland}	
	
\author{F. Carbone}
	\affiliation{Institute of Physics and Lausanne Center for Ultrafast Science (LACUS), \'Ecole Polytechnique F\'{e}d\'{e}rale de Lausanne, CH-1015 Lausanne, Switzerland}

\date{\today}

\begin{abstract}
Recent ultrafast magnetic-sensitive measurements [Phys. Rev. B \textbf{92}, 184429 (2015) and Phys. Rev. B \textbf{96}, 184414 (2017)] have revealed a delayed melting of the long-range cycloid spin-order in TbMnO$_3$ following photoexcitation across the fundamental Mott-Hubbard gap. The microscopic mechanism behind this slow transfer of energy from the photoexcited carriers to the spin degrees of freedom is still elusive and not understood. Here, we address this problem by combining spectroscopic ellipsometry, ultrafast broadband optical spectroscopy and \textit{ab initio} calculations. Upon photoexcitation, we observe the emergence of a complex collective response, which is due to high-energy coherent optical phonons coupled to the out-of-equilibrium charge density. This response precedes the magnetic order melting and is interpreted as the fingerprint of the formation of anti-Jahn Teller polarons. We propose that the charge localization in a long-lived self-trapped state hinders the emission of magnons and other spin-flip mechanisms, causing the energy transfer from the charge to the spin system to be mediated by the reorganization of the lattice. Furthermore, we provide evidence for the coherent excitation of a phonon mode associated with the ferroelectric phase transition.
\end{abstract}


\maketitle

\section{I. Introduction}

In the last years, increasing attention has been devoted to the study of type-II multiferroics, with the aim of using electric fields to switch magnetic order, or magnetic fields to switch the spontaneous electric polarization \cite{cheong2007multiferroics}. One of the ultimate goals in this research area involves the ability to achieve dynamic switching on ultrafast timescales by means of tailored laser pulses. To better understand the dynamics of magnetic order, several time-resolved techniques that are directly sensitive to long-range spin order are currently being explored. For example, time-resolved cryo-Lorentz transmission electron microscopy holds huge promise for imaging the dynamical evolution of magnetic correlations in real space with nanometer spatial resolution \cite{park20104d, rajeswari2015filming}. This approach has recently reached a temporal resolution of 700 fs, thus allowing for the observation of fast spin-related phenomena occurring below the 50 ps timescale \cite{ropers}. An alternative method is represented by time-resolved resonant elastic x-ray scattering (trREXS), which is currently capable of a time resolution of 100 fs in slicing facilities at third-generation synchrotron sources and in free-electron lasers \cite{fink2013resonant, ref:kubacka, ref:johnson, ref:johnsonCuO, langner2015ultrafast, bothschafter_TMO}. Within this approach, after the photoexcitation with an ultrashort laser pulse, a soft x-ray probe tuned to a resonance can provide sensitivity to spin, orbital, and charge order in combination with element selectivity.

\begin{figure}[tb]
	\centering
	\includegraphics[width=0.7\columnwidth]{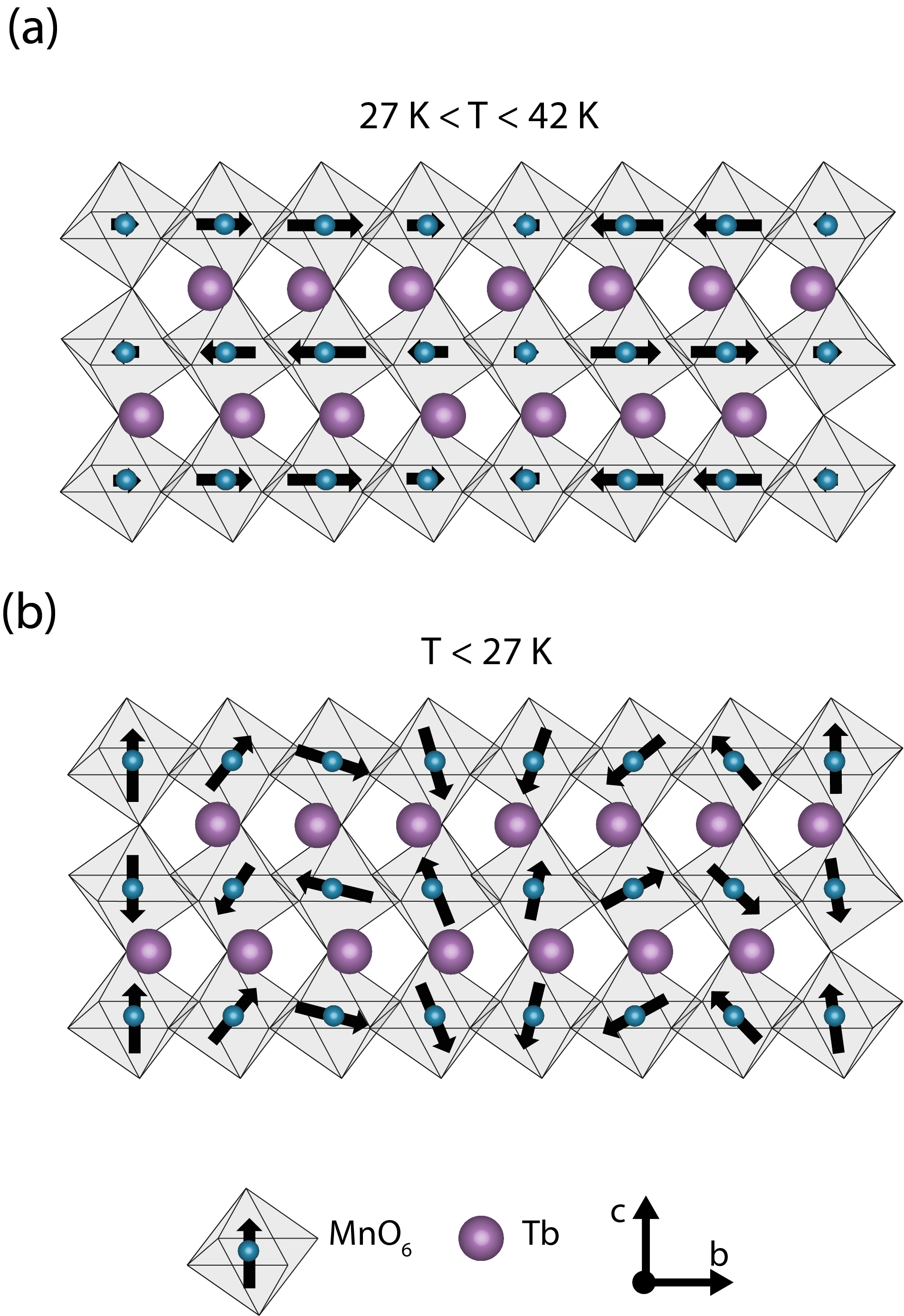}
	\caption{Schematic representation of spin order in TbMnO$_3$ (a) in the paraelectric SDW state ($\mathrm{T_{N2}}$ $<$ T $<$ $\mathrm{T_{N1}}$) and (b) in the spin cycloid phase (T $<$ $\mathrm{T_{N2}}$). Here $\mathrm{T_{N1}}$ = 42 K, $\mathrm{T_{N2}}$ = 27 K. Mn atoms are reported in blue, Tb atoms in violet and O atoms are omitted at the vertices of the MnO$_6$ octahedra. Ellipticity in (b) has been neglected.}
	\label{fig:TMO_Spin}
\end{figure} 

Within the class of type-II multiferroics, one of the most studied systems is TbMnO$_3$, which crystallizes in a GdFeO$_3$-distorted structure (space group \textit{Pbnm}). With regard to its magnetic properties, many competing exchange paths exist between nearest neighbour and next nearest neighbour Mn ions, which lead to frustration of the magnetic moments on the Mn$^{3+}$ sites \cite{ref:kimura_TMO}. This magnetic frustration is at the origin of the exotic long-range spin ordering patterns that characterize TbMnO$_3$ \cite{ref:quezel_TMO, ref:kenzelmann_TMO, ref:senff_TMO, ref:wilkins_TMO, ref:walker_TMO, ref:lovesey_TMO}. At room temperature this material is paramagnetic. When the temperature is decreased below 100 K, short-range spin correlations start to develop and a magnetic and structural fluctuating regime dominates over a wide temperature range. The first magnetic phase transition is observed at $\mathrm{T_{N1}}$ = 42 K with the establishment of a paraelectric spin-density wave (SDW) state, in which a sinusoidal antiferromagnetic spin structure forms along the \textit{b}-axis (Fig. \ref{fig:TMO_Spin}(a)). A second magnetic phase transition occurs around $\mathrm{T_{N2}}$ = 27 K, at which the spins on the Mn$^{3+}$ sites order in a spin cycloid in the \textit{bc} plane (Fig. \ref{fig:TMO_Spin}(b)). An additional transition related to the 4\textit{f} spins of the Tb$^{3+}$ ions is finally observed at 7 K \cite{ref:kenzelmann_TMO, ref:kimura_nature}. The cycloid spin ordering of the Mn$^{3+}$ sites below $\mathrm{T_{N2}}$ creates a spontaneous ferroelectric polarization along the \textit{c}-axis due to the combined effects of the inverse Dzyaloshinskii-Moriya interaction and of the Tb$^{3+}$ ionic displacement \cite{ref:kimura_nature, ref:goto, ref:kenzelmann_TMO, ref:walker_TMO}.

By using trREXS upon resonant THz excitation of an electromagnon in the cycloid spin-ordered phase, it has been recently demonstrated that a coherent dynamics of the magnetic structure can be driven \cite{ref:kubacka}. On the other hand, optical excitation in the near-infrared/visible range has been shown to lead to a delayed melting of the long-range magnetic order on a time scale of $\sim$20 ps \cite{ref:johnson}. Time-resolved measurements of magnetic x-ray scattering have suggested that the loss of magnetic order and the associated Mn 3\textit{d} orbital reconstruction follows a quasi-adiabatic pathway from the cycloidal phase through an intermediate sinusoidal phase, similar to what is seen when slowly increasing the temperature \cite{bothschafter_TMO}. Unlike the situation seen in thermal equilibirum, the wavevector of the spin ordering in the intermediate sinusoidal phase remains at a value equivalent to the cycloidal ordering wavevector. A direct coupling between magnetic and orbital orders was also found by monitoring the orbital reconstruction induced by the spin ordering \cite{bothschafter_TMO}. Although these studies shed light on various mechanisms occurring in TbMnO$_3$ upon photoexcitation, a complete microscopic explanation for the delayed demagnetization process is still lacking. Such slow dynamics imply a delayed transfer of energy from the photoexcited carriers to the spin system, which is at odds with the behavior observed in other correlated oxides with long-range spin order \cite{ref:forst_manganite, ref:johnsonCuO}. As such, the dynamic bottleneck may be due to the lattice or to the spin system itself, or to a combination of both.

A promising route to elucidate the energy pathway and the possible involvement of other degrees of freedom in the melting process is to perform ultrafast optical spectroscopy with a combination of broad wavelength range and high temporal resolution \cite{mann2015probing, mann2016probing, baldini_MgB2, borroni2017coherent}. A broad wavelength range is crucial when large shifts of the optical spectral weight (SW) are induced by magnetic ordering phenomena. Photons in the optical range can access magnetic and other complex ordering effects, as changes of the symmetry in a system upon ordering are often reflected in the optical properties of the material. This mechanism is at play in the insulating perovskite manganites and manifests on the \textit{d}-\textit{d} intersite transitions of the optical spectrum \cite{ref:kovaleva_prl}. With a time resolution below 50 fs, one can reveal the timescales on which the excited electronic states redistribute their excess energy, unveiling the fingerprint, if any, of specific collective modes associated with the ultrafast rearrangement of the lattice or spin subsystems. In this work, we combine steady-state spectroscopic ellipsometry (SE), ultrafast broadband optical spectroscopy with a time resolution of 45 fs and \textit{ab initio} calculations to elucidate the hierarchy of phenomena following the photoexcitation of TbMnO$_3$ across the fundamental Mott gap and leading to the delayed melting of the cycloid spin-order. Our results are suggestive of a scenario in which anti-Jahn-Teller (JT) polarons are formed upon photoexcitation, leading to the emergence of a fast coherent lattice response. We propose that the charge localization in a long-lived self-trapped state hinders the emission of magnons and other spin-flip mechanisms, causing the energy transfer from the charge to the spin system to be mediated by the reorganization of the lattice.

The paper is organized as follows: In Sec. II we report the steady-state optical properties of the material using SE and we perform \textit{ab initio} calculations to assign the features in the optical spectra. Next, in Sec. III, we show the out-of-equilibrium experiments and describe the spectro-temporal signatures for the melting of the long-range spin order. We conclude with Sec. IV, offering a complete picture of the ultrafast dynamics following photoexcitation of TbMnO$_3$ across the Mott gap.

\section{II. Steady-state optical properties}
\label{Optics_TMO}

\subsection{A. Spectroscopic ellipsometry}

Here we present temperature dependent SE measurements of a detwinned (010)-oriented single crystal of TbMnO$_3$ in the spectral range between 0.50 eV and 6.00 eV. Details on the sample preparation and the SE experiments are given in the Supplementary Materials (SM). Figures \ref{fig:Conductivity}(a-d) show the temperature dependence of the real- ($\sigma_1$) and imaginary ($\sigma_2$) parts of the optical conductivity, measured  with light polarized parallel to the \textit{a}- (\textbf{E} $\parallel$ a) and \textit{c}-axis (\textbf{E} $\parallel$ c) of the crystal. We observe a marked anisotropy between the $a$- and $c$-axis response, which confirms the complete detwinning of our TbMnO$_3$ single crystal. The $\sigma_{1a}$ spectrum is featureless in the near-infrared region and is dominated by two broad absorption bands, centred around 2.45 eV and 5.15 eV at all measured temperatures. As the temperature is increased from 8 K to 300 K, a number of smaller spectral features emerge on top of the optical band at 2.45 eV. In contrast, the $\sigma_{1c}$($\omega$) spectrum is characterized by a  strongly increasing conductivity with frequency, particularly between 2.00 and 3.00 eV, with complex features appearing at 3.80 eV and 6.00 eV that evolve with temperature. The latter consists of an optical band centered around 4.50 eV at 8 K, which progressively loses SW for increasing temperature and splits into two well distinguished features at 3.90 eV and 5.15 eV above 25 K. The observed anisotropic response is in agreement with that reported in a previous study on detwinned TbMnO$_3$ single crystals \cite{ref:bastjan}.

\begin{figure}[tb]
\centering
\includegraphics[width=\columnwidth]{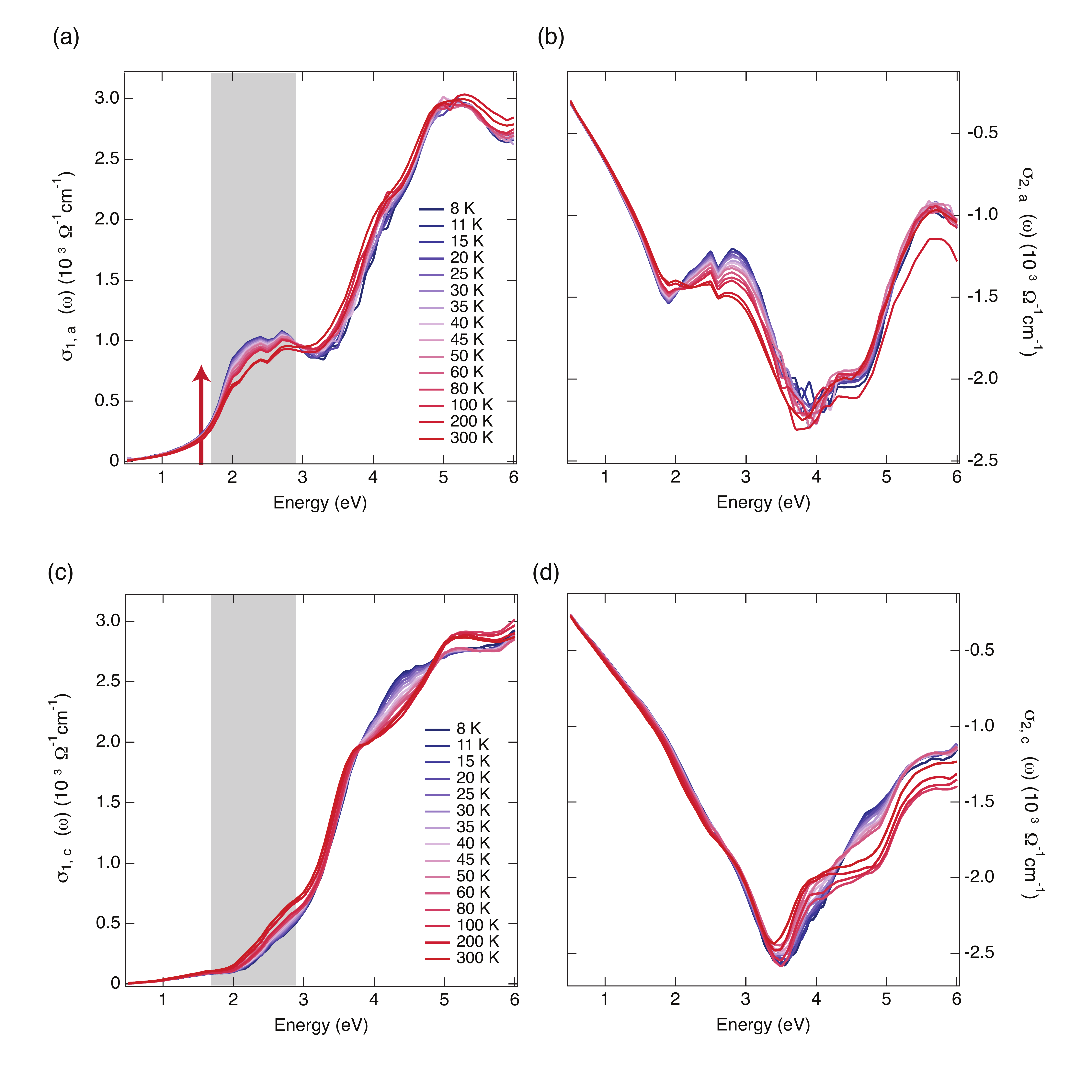}
\caption{Temperature dependence of the (a,c) real $\sigma_1$ and (b,d) imaginary $\sigma_2$ parts of the complex optical conductivity of TbMnO$_3$; (a,b) refer to the optical response for light polarized along the \textit{a}-axis, (c,d) for light polarized along the \textit{c}-axis. The red arrow indicates the pump photoexcitation at 1.55 eV in the nonequilibrium experiment. The grey shaded area represents the spectral region monitored by the probe along the two different axes.}
\label{fig:Conductivity}
\end{figure}

\begin{figure}[tb]
	\centering
	\includegraphics[width=\columnwidth]{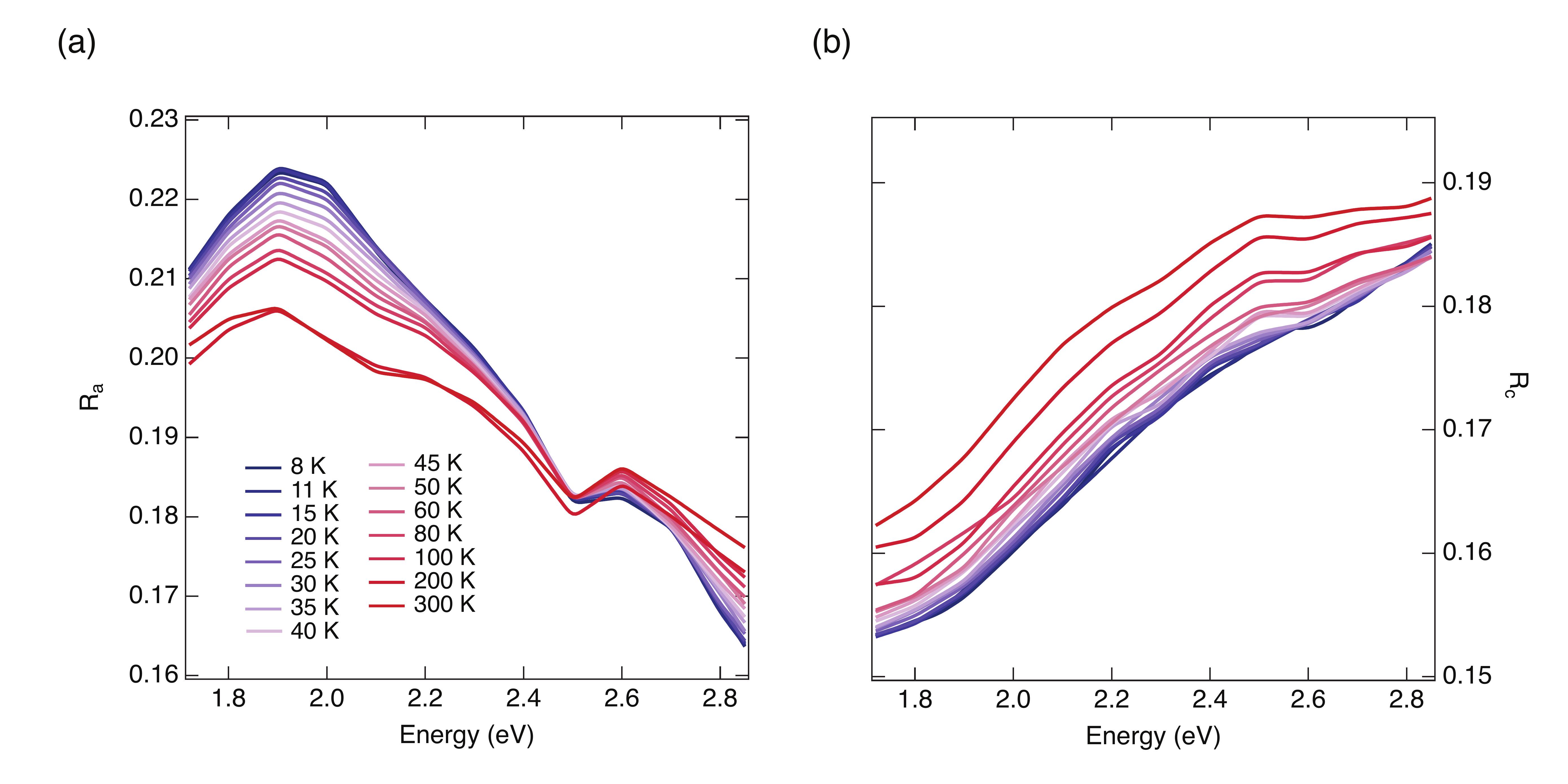}
	\caption{Temperature dependence of the (a) \textit{a}-axis (b) \textit{c}-axis reflectivity of TbMnO$_3$.}
	\label{fig:Reflectivity}
\end{figure}

From our SE measurements, we calculate the anisotropic reflectivity spectra (R) under equilibrium conditions, since our ultrafast study involves the measurements of the transient reflectivity ($\Delta$R/R) of TbMnO$_3$. The spectra are shown as a function of temperature in Figs. \ref{fig:Reflectivity}(a,b) along the \textit{a}- and \textit{c}-axis, respectively. The displayed range is limited to the one covered by the broadband probe in the nonequilibrium experiment. The spectrum of R along the \textit{a}-axis consists of a pronounced feature centred around 1.90 eV with a fine structure near its maximum. A second feature emerges around 2.60 eV, separated from the former by a dip in the reflectivity. Along the \textit{c}-axis, the reflectivity monotonically increases with higher energies and shows a rich structure in the 2.40 - 2.80 eV spectral region. Consistent with the SW shifts identified in the previous discussion, we observe an opposite trend of R as a function of temperature for the two different axes. While R$_a$ undergoes a prominent drop with increasing temperature, R$_c$ increases over the whole spectral range. This aspect enters into the interpretation of the pump-probe data presented in Sec. III.

\subsection{B. Assignment of the optical features}

To assign the observed structures, we rely on the results reported in literature on orthorhombic manganites \cite{ref:okimoto_TMO, ref:takenaka_TMO, ref:tobe_TMO, ref:quijada_TMO, ref:kovaleva_prl, ref:bastjan, ref:kovaleva_prb, ref:nucara, ref:kimeffect, ref:lawler_TMO, rusydi, ref:moskvin_TMO, ref:kovaleva_orbital} and perform \textit{ab initio} calculations based on Density-Functional Theory (DFT). First, we focus on the main features characterizing the optical spectrum. Previous work attempted to clarify the origin of the different absorption bands by identifying possible signatures of the magnetic ordering on the optical response. Particular attention was devoted to LaMnO$_3$, which retains the lowest degree of structural distortion from the ideal perovskite. Early studies assigned both the low-energy and high-energy bands to \textit{p}-\textit{d} charge-transfer (CT) excitations \cite{ref:okimoto_TMO, ref:takenaka_TMO, ref:tobe_TMO} or associated the low-energy feature with JT orbiton excitations or \textit{d}-\textit{d} crystal field transitions \cite{ref:allen_perebeinos}. Later on, the validity of these assignments was questioned by the observation of pronounced rearrangements of the SW between the two bands close to $\mathrm{T_N}$ = 140 K  \cite{ref:quijada_TMO, ref:kovaleva_prl, ref:kovaleva_prb}. It was concluded that the broad low-energy band in the \textit{ab}-plane response of LaMnO$_3$ originates from intersite Mn$^{3+}$ \textit{d}-\textit{d} CT transitions (\textit{i.e.} $d_i^4$ $d_j^4$ $\rightarrow$ $d_i^3$ $d_j^5$ between neighbouring \textit{i} and \textit{j} Mn ions), preserving the electronic spin state. It is therefore a high-spin (HS) transition that is favoured in the antiferromagnetic phase. The \textit{c}-axis low-energy optical response is instead governed by the CT between Hund states of neighbour Mn$^{3+}$ ions with antiparallel spin. Thus, it involves a low-spin (LS) transition. In contrast, the more isotropic high-energy band was attributed to the manifold of CT transitions from O 2\textit{p} to Mn 3\textit{d} levels \cite{rusydi}. Such an assignment has a profound consequence for the interpretation of the perovskite manganites electronic structure, as it implies that these materials are Mott-Hubbard insulators and not \textit{p}-\textit{d} CT insulators \cite{ref:zaanen}. The HS band should be then interpreted as an intersite \textit{d}-\textit{d} CT transition across the Mott gap, between the LHB and the UHB. The same conclusion was drawn by calculations based on the orbitally-degenerate Hubbard model \cite{ref:lee2005optical}. In addition, a clear manifestation of the \textit{d}-\textit{d} CT origin of the low-energy band comes from the resonant enhancement at 2.00 eV of the $\mathrm{B_{2g}}$ breathing mode in spontaneous Raman scattering, as this mode strongly modulates the intersite \textit{d}-\textit{d} CT \cite{ref:kruger_TMO}.

\begin{figure}[tb]
	\centering
	\includegraphics[width=\columnwidth]{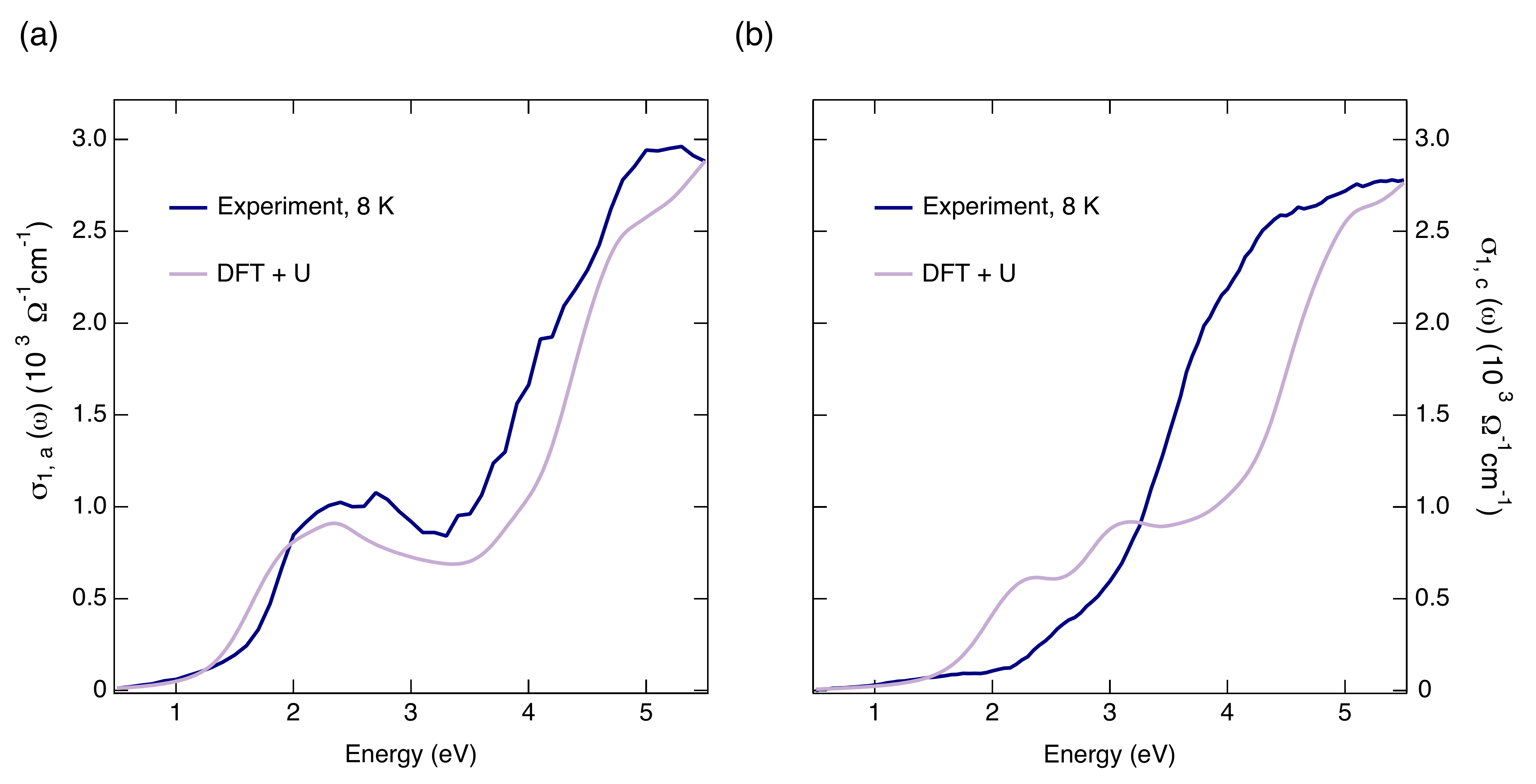}
	\caption{Comparison between $\sigma_1$($\omega$) of TbMnO$_3$ measured at 8 K (blue curve) and computed from DFT + U calculations (violet curve): (a) $a$-axis response; (b) $c$-axis response.}
	\label{fig:DFT}
\end{figure}

Here, we confirm the latter assignments by employing \textit{ab initio} calculations of the electronic structure and the optical properties of TbMnO$_3$.  Our calculations were carried out using DFT via Wien2k code \cite{wien2k}. The exchange-correlation potential were treated using GGA + U, where U was set at 3 and 6 eV for Mn 3$d$ and Tb 4$f$ orbitals, respectively. The computational details are reported in the SM. The calculated equilibrium optical responses along the $a$- and $c$-axes are shown in Fig. \ref{fig:DFT}(a,b) together with the experimental data measured at 8 K. For the $a$-axis response we see a relatively good agreement between our calcualtions and the experiment. For the $c$-axis response the agreement is mostly qualitative but does capture the rough position and magnitude of the increase in absorption with increasing energy. Based on the calculated electronic density of states, the low-energy feature is identified with intersite $d$-$d$ transitions while the higher energy feature is identified with CT from O 2$p$ to Mn 3$d$ levels.

These combined experimental-theoretical efforts clarify the coarse features of the optical spectrum, but the origin of the fine structure on the HS band still remains largely unexplained. Recently, several authors have attempted to better understand this fine structure in LaMnO$_3$. Initially, it was attributed to Mn$^{3+}$ \textit{d}-\textit{d} crystal field transitions split by the JT effect \cite{ref:lawler_TMO}. A complete and detailed study assigned a dual nature (\textit{d}-\textit{d} and \textit{p}-\textit{d}) to the fundamental optical gap, since this is also characterized by forbidden/weakly-allowed \textit{p}-\textit{d} transitions that act as a precursor to the strong dipole-allowed \textit{p}-\textit{d} CT transition at higher energies \cite{ref:moskvin_TMO}. Thus, the fine structure reflects these \textit{p}-\textit{d} transitions overlapped to the \textit{d}-\textit{d} HS absorption band, giving rise to an intermediate regime between the \textit{p}-\textit{d} CT and the Mott-Hubbard insulator in the Zaanen-Sawatzky-Allen scheme \cite{ref:zaanen}. A more recent interpretation associates the fine structure with quantum rotor orbital excitations for the $e_g$ electron of Mn$^{3+}$ ions, disturbed by the lattice anharmonicity \cite{ref:kovaleva_orbital}. To date there is no clear consensus on the origin of the spectral fine structure. In contrast to LaMnO$_3$, TbMnO$_3$ retains such a fine structure of the \textit{d}-\textit{d} HS band even at very low temperatures, as evident in the high-resolution spectrum of $\sigma_{1,a}$($\omega$) in Fig. S1(a). Moreover, the centre of mass of the HS optical band in TbMnO$_3$ is shifted towards higher energies than the corresponding one in LaMnO$_3$. This effect has been explained by assuming that the JT distortion increases for smaller R ion sizes, thus enhancing the value of the \textit{d}-\textit{d} CT energy \cite{ref:moskvin_TMO}.

\section{III. Ultrafast broadband optical spectroscopy}

In this Section, we present an extensive study of the ultrafast dynamics occurring in our detwinned (010)-oriented single crystal of TbMnO$_3$. Details on our experimental setup are given in Ref. \cite{baldini2016versatile} and in the SM. Briefly, we drive the system out-of-equilibrium using an ultrashort 1.55 eV pump pulse, which mimics the experimental conditions of previous trREXS studies \cite{ref:johnson, bothschafter_TMO}. The pump beam is polarized along the \textit{a}-axis and it mainly promotes intersite \textit{d}-\textit{d} transitions \cite{ref:kovaleva_prl} (red arrow in Fig. 2(a)). The possibility that the excitation of \textit{p}-\textit{d} CT transitions could lead to the magnetic dynamics was ruled out via trREXS experiments by observing the persistence of the average scattering cross-section intensity. Although this argument is not sufficient to exclude that tails of the \textit{p}-\textit{d} CT excitation could contribute to other channels probed in the optical range, in the following we will assume that the main effect of photoexcitation is to promote intersite \textit{d}-\textit{d} transitions. After the interaction with the pump pulse, the ultrafast variation of the material reflectivity ($\Delta$R/R) is monitored over a broad visible range, extending between 1.72 eV and 2.85 eV, along both the \textit{a}- and \textit{c}-axis. This is highlighted by a grey shaded area in Fig. \ref{fig:Conductivity}(a,c). As described in Sec. II, this spectral region involves the intersite HS and LS \textit{d}-\textit{d} transitions along the \textit{a}- and \textit{c}-axes, respectively \cite{ref:kovaleva_prl}. As the optical features depend on the establishment or disappearance of the magnetic order in the crystal, they can represent suitable observables for extracting valuable information on the optical properties of the magnetic phase with high time resolution. Indeed, although all-optical pump-probe methods do not offer a direct measure of the magnetic order dynamics, the change in the optical properties of the magnetic phase still provides a qualitative estimate of the timescale involved in the magnetic order melting \cite{bothschafter_TMO}. 

\begin{figure}[tb]
\centering
\includegraphics[width=\columnwidth]{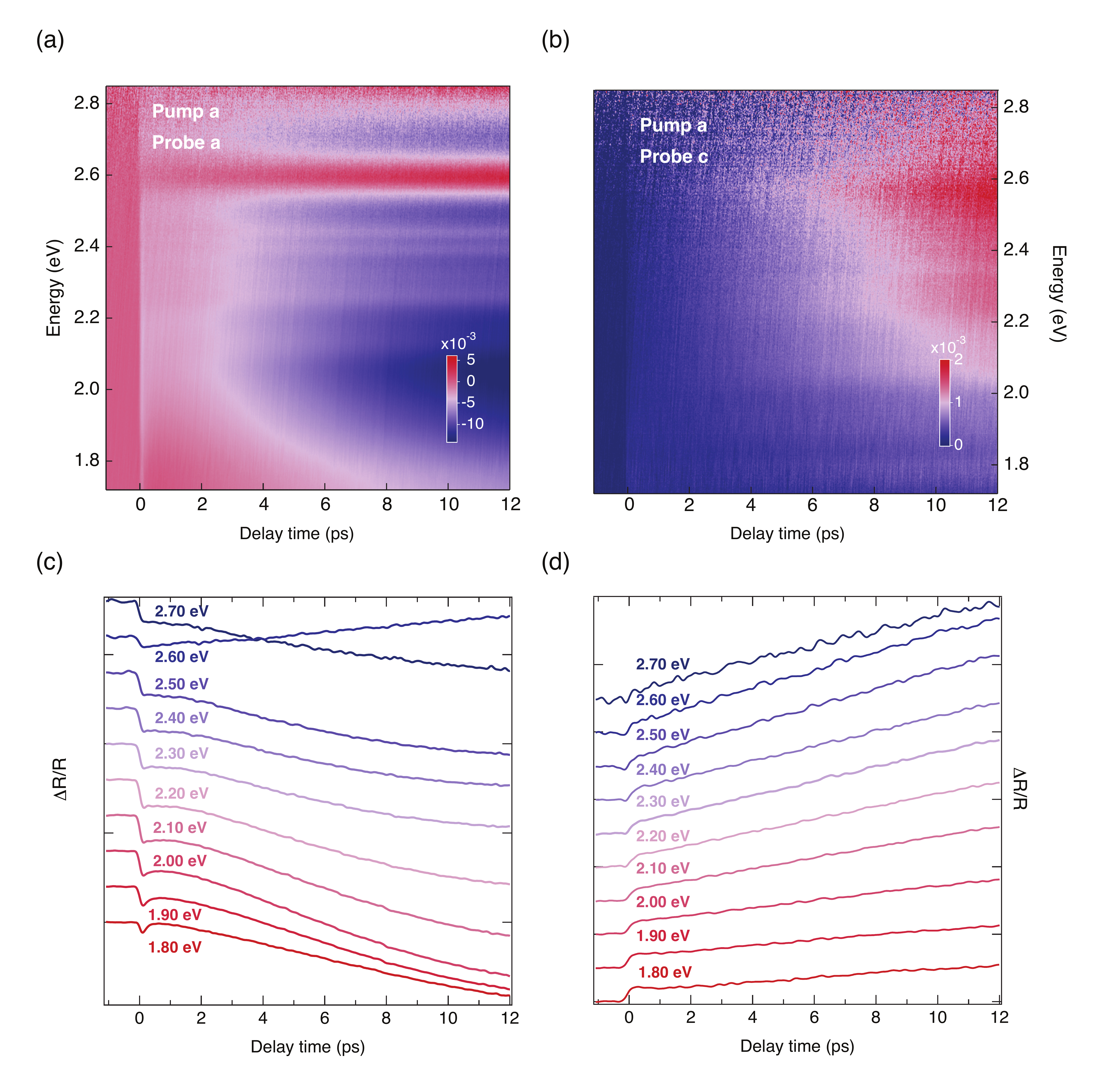}
\caption{(a,b) Colour-coded maps of $\Delta$R/R at 8 K with \textit{a}-axis pump polarization and (a) \textit{a}-axis, (c) \textit{c}-axis probe polarization. The pump photon energy is 1.55 eV and the absorbed pump fluence is 4.4 mJ/cm$^2$. (c,d) Temporal traces at specific probe photon energies of the respective $\Delta$R/R maps. Each temporal trace results from the integration over 0.10 eV around the indicated probe photon energy.}
\label{fig:Spectrum_aa_ac_TMO}
\end{figure}

\subsection{A. Low-temperature $\Delta$R/R}

Figures \ref{fig:Spectrum_aa_ac_TMO}(a,b) display the color maps of $\Delta$R/R as a function of the probe photon energy and of the time delay between pump and probe at 8 K, for a probe polarization along the \textit{a}- and \textit{c}-axes. In both cases, the pump polarization lies along the \textit{a}-axis and the absorbed pump fluence is 4.4 mJ/cm$^2$. Importantly, in this set of measurements, the dynamics are detected up to 14 ps, while sampling with a time step of $\sim$26 fs. The \textit{a}-axis $\Delta$R/R consists of a negative signal in correspondence to the HS intersite \textit{d}-\textit{d} transition, which increases its absolute weight over time. The spectral shape of $\Delta$R/R also acquires more features over time, displaying a series of satellites. Simultaneously, a very sharp feature emerges around 2.60 eV and changes the sign of $\Delta$R/R to positive. In contrast, the \textit{c}-axis response undergoes a remarkably different behavior. It is dominated by a positive signal, which increases for longer time delays and shows a maximum close to 2.60 eV. 

More insight on the temporal dynamics can be gained by selecting different time traces at representative photon energies. These are shown in Figs. \ref{fig:Spectrum_aa_ac_TMO}(c,d). We first consider the dynamics seen for probe light polarized along the crystal $a$-axis, as shown in Fig. \ref{fig:Spectrum_aa_ac_TMO}(c). Here we see that for all probe photon energies at short times there is a sudden decrease in reflectivity which reaches a local minimum at approximately 190 fs.  This time scale is significantly longer than the 45 fs response time of the apparatus \cite{baldini2016versatile}.  After this prompt decrease, there is a fast partial recovery of the reflectivity that is characterized by a time scale of several hundred femtoseconds, a recovery that is particularly evident at lower photon energies ($e.g.$ 1.80 eV).  After this partial recovery, the reflectivity changes more slowly.  For most photon energies this change is a slow decrease in reflectivity that continues beyond the time window of our measurement. At 2.60 eV, however, the reflectivity instead increases. Here, a negative-to-positive crossover clearly develops after 3.8 ps, as also evident from the map of Fig. \ref{fig:Spectrum_aa_ac_TMO}(a). Probing in a broad spectral range allows us to go beyond the results obtained by previous two-color pump-probe experiments on TbMnO$_3$, in which the fast components in the \textit{a}-axis response was not detected \cite{ref:handayani}.

Along the \textit{c}-axis, the temporal traces show a markedly different behavior. The rise time of the signal is $\sim$400 fs and all temporal traces undergo a long increase of the response amplitude over time. Here the distinction between a short-lived component and a delayed one is not pronounced and no changes of sign appear over time.

\begin{figure}[t]
\centering
\includegraphics[width=\columnwidth]{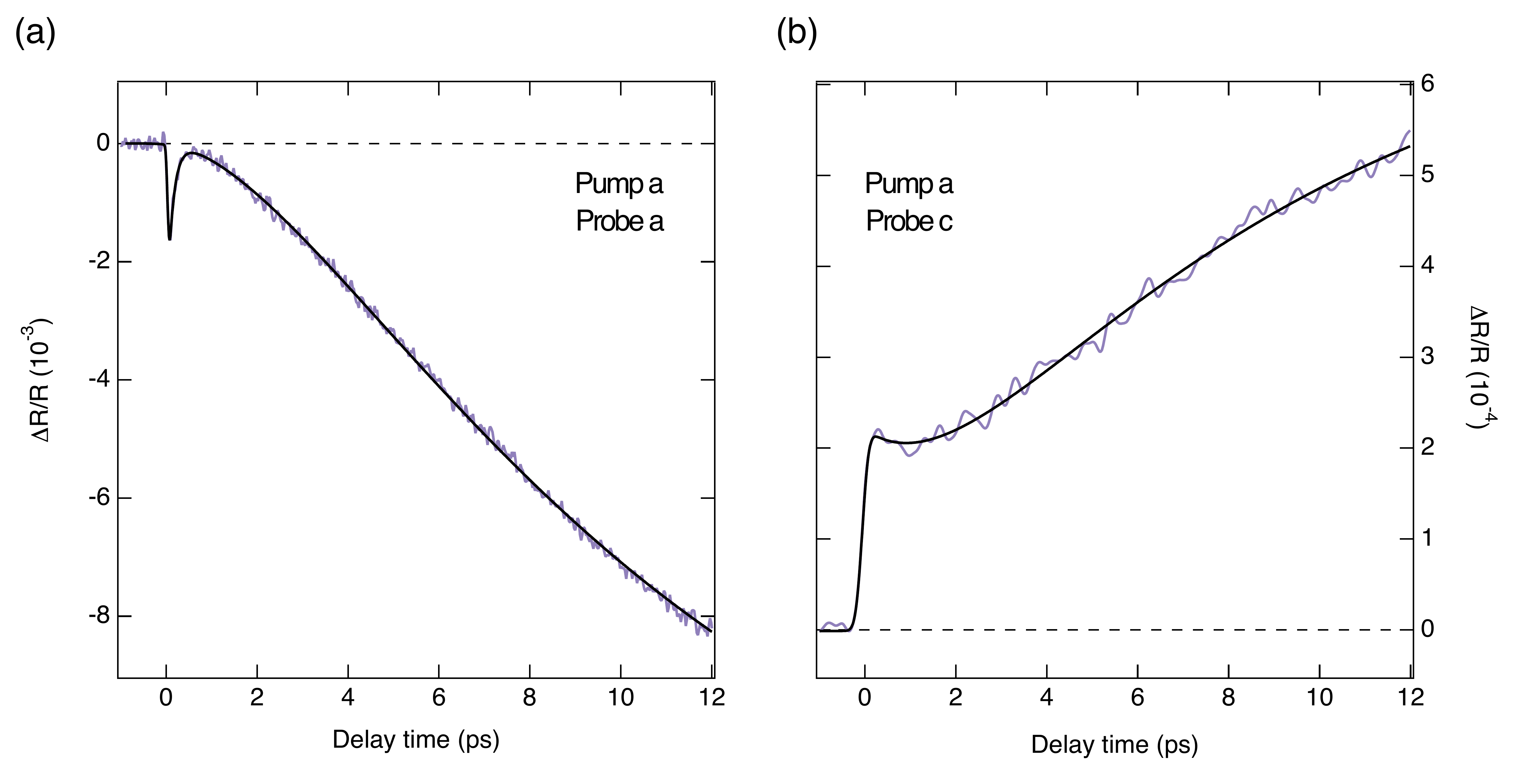}
\caption{Temporal traces of the $\Delta$R/R response at 8 K along the (a) \textit{a}- and (b) \textit{c}-axis, resulting from the integration over 0.10 eV around the probe photon energy of 1.80 eV.  In both cases, the pump polarization lies along the \textit{a}-axis and the absorbed pump fluence is estimated around 4.4 mJ/cm$^2$. The black lines represent the results of the fit based on our model function.}
\label{fig:Fit_TMO}
\end{figure}

To get more quantitative information on the timescales governing the dynamics, we perform a fit of representative temporal traces at 8 K. For this purpose, we select the temporal traces around 1.80 eV, since the short-lived component becomes more apparent in this region. The traces along the \textit{a}- and \textit{c}-axis are respectively shown as violet curves in Figs. \ref{fig:Fit_TMO}(a,b). We find that the simplest model that captures the dynamics consists of three exponential functions (with time constants $\tau_{1}$, $\tau_{2}$ and $\tau_{m}$) convolved with a Gaussian accounting for the temporal shape of the pump pulse. The shortest lived component appears immediately after the pump photoexcitation, whereas the other two display a slow rise. More details about the fitting procedure are reported in the SM.

While all three exponential functions are necessary for fitting the \textit{a}-axis response, only two are needed to reproduce the \textit{c}-axis dynamics. The results of the fit are displayed in Figs. \ref{fig:Fit_TMO}(a,b) as black lines superimposed on the original data. The timescales retrieved along the \textit{a}-axis are $\tau_{1}$ = 120 $\pm$ 10 fs, $\tau_{2}$ = 3 $\pm$ 0.7 ps and $\tau_{m}$ = 9 $\pm$ 1.5 ps, while those along the \textit{c}-axis are $\tau_{1}$ = 2 $\pm$ 0.5 ps and $\tau_{m}$ = 9 $\pm$ 0.5 ps. A larger uncertainty is expected on $\tau_{m}$ due to the limited temporal window of $\sim$12 ps probed in our experiment. Between the two axes, we notice a strong mismatch in the time constant $\tau_{1}$ and a close correspondence in the time constant $\tau_{m}$. Considering the \textit{a}-axis response alone, the emerging picture of the temporal dynamics reconciles the results of separate two-color pump-probe experiments performed in the past on LaMnO$_3$ and TbMnO$_3$ \cite{ref:wall_manganite, ref:handayani}. Analysis of a high time-resolution experiment on LaMnO$_3$ attributed a short-lived component similar in time-scale to our $\tau_1$ to electron thermalization, while the intermediate time-scale $\tau_2$ was interpreted as electron-phonon relaxation \cite{ref:wall_manganite}. Another experiment on TbMnO$_3$ had significantly worse time resolution and considered only the slowest component of the time-resolved changes along both axes (corresponding to our $\tau_m$), interpreting this as the time scale for melting of magnetic order \cite{ref:handayani}. Despite observing also the faster component along the \textit{c}-axis, the latter experiment explicitly neglected it due to the difficulty in fitting the data. Using our model function, we find this component to decay with a time constant of $\tau_{1}$ = 2 ps, which is one order of magnitude larger than expected from a typical electron-electron thermalization timescale.

\begin{figure}[tb]
\centering
\includegraphics[width=\columnwidth]{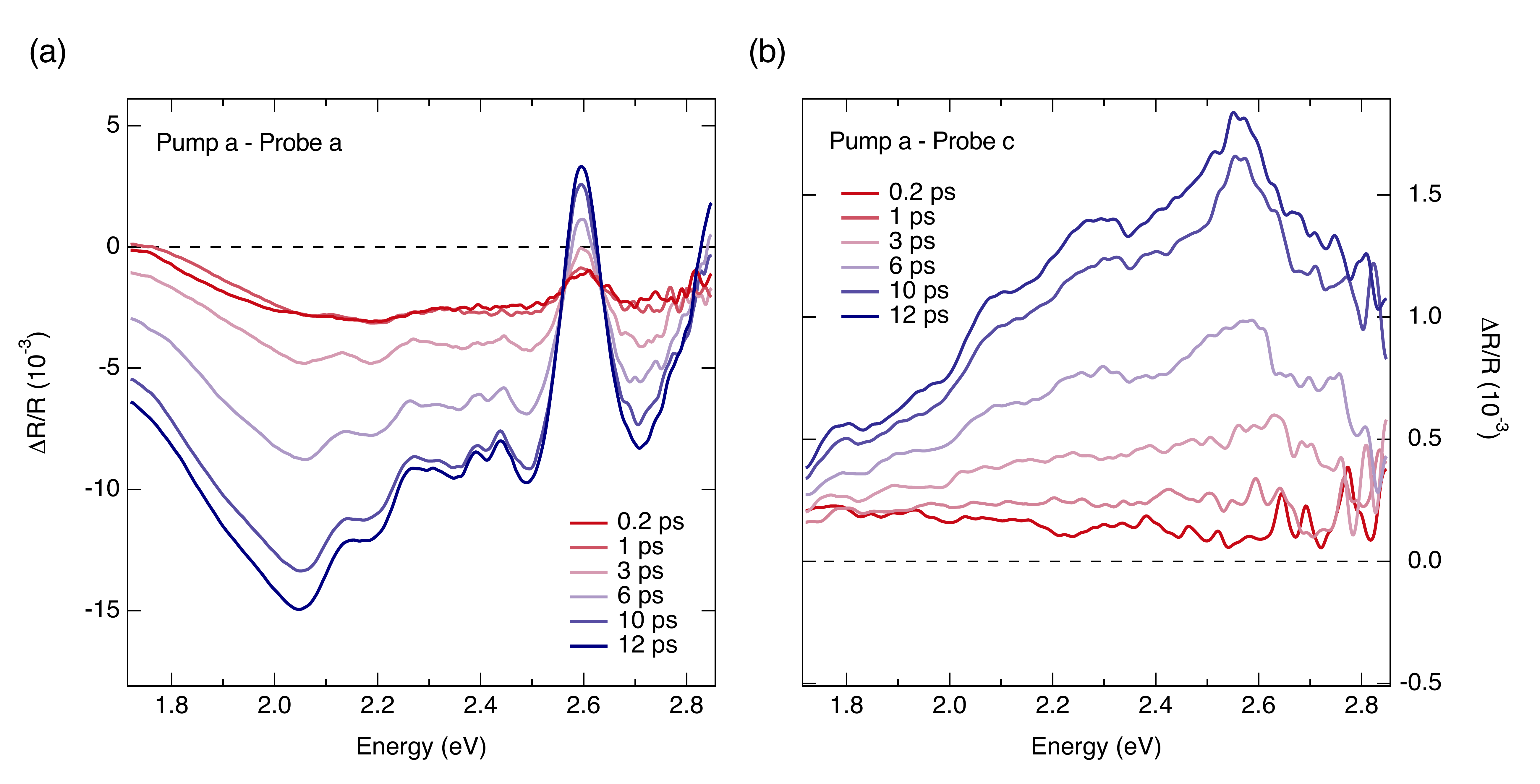}
\caption{Transient spectrum of $\Delta$R/R at different delay times for a probe polarization set along (a) the \textit{a}-axis, (b) the \textit{c}-axis. The pump photon energy is 1.55 eV and the absorbed pump fluence is 4.4 mJ/cm$^2$.}
\label{fig:TransientSpectrum_TMO}
\end{figure}

As an alternative way of examining the data, in Fig. \ref{fig:TransientSpectrum_TMO}(a,b) we show the spectral evolution of $\Delta$R/R at different time delays along the \textit{a}- and \textit{c}-axes, respectively. In Fig. \ref{fig:TransientSpectrum_TMO}(a), the \textit{a}-axis $\Delta$R/R spectrum for early time delays is almost featureless and shows only a broad structure around 2.60 eV. As time evolves, a fine peak-dip structure clearly emerges in the spectrum, with shoulders covering the range from 2.05 eV to 2.50 eV. These features coincide with the fine structure superimposed on the \textit{d}-\textit{d} HS absorption band, which can also be observed in the steady-state reflectivity spectrum of Fig. \ref{fig:Reflectivity}(a). At 2.60 eV the former broad feature sharpens and increases its weight until the sign of the response is preserved. Above 2.85 eV the $\Delta$R/R response changes sign and becomes positive, as evidenced by the gradual redshift of the zero-crossing energy at the edge of the probed spectrum. In Fig. \ref{fig:TransientSpectrum_TMO}(b), the \textit{c}-axis $\Delta$R/R spectrum for early time delays appears as a featureless background, while for long time delays some fine structure also arises. Remarkably, these features are not apparent under steady-state conditions (Fig. \ref{fig:Reflectivity}(b)) and cannot be related to a leakage of the \textit{a}-axis response, since their energies are well distinct from the ones of the \textit{a}-axis fine structure. Thus, we conclude that our pump-probe experiment provides a higher contrast to resolve elementary excitations that are hidden in the equilibrium spectra. However, at this stage, no insightful information can be retrieved on the origin of this fine structure characterizing the optical spectrum of TbMnO$_3$.

From this preliminary analysis, we can conclude that the photoexcitation of TbMnO$_3$ along the \textit{a}-axis with a pump photon energy of 1.55 eV leads to an overall reduction of the reflectivity along the \textit{a}-axis, accompanied by an increase of the reflectivity along the \textit{c}-axis. This anisotropic behavior goes beyond the results of previous two-color pump-probe measurements performed at different temperatures \cite{ref:handayani}. In these experiments, the pump pulse was set at 3.00 eV, thus at the edge between the intersite \textit{d}-\textit{d} transitions and the tail of the CT transition, while the probe pulse at 1.50 eV was monitoring the lowest tail of the HS and LS bands in the material. As a consequence, the probe was not capable of revealing the details of the ultrafast electronic response at early time delays, nor of providing spectrally resolved information across the whole spectral region of the intersite \textit{d}-\textit{d} excitations. In the following, we demonstrate instead that our approach bridges the gap between the conclusions drawn by ultrafast two-color optical spectroscopy and trREXS experiments, thus providing a unified picture of the nonequilibrium dynamics of TbMnO$_3$ triggered by a near-infrared pump pulse.

\subsection{B. Temperature dependence}

As highlighted in Sec. II, the spectral region of the \textit{a}-axis HS optical band is indirectly sensitive to the magnetic correlations in the insulating perovskite manganites \cite{ref:kovaleva_prl}. We can then expect that measurements of optical properties in this frequency range can give us some information on the dynamics associated with magnetic order changes. A useful quantity that can be extracted from the nonequilibrium experiment is the transient complex optical conductivity $\Delta\sigma$ = $\Delta\sigma_1$ + i $\Delta\sigma_2$. This is estimated without the need of a Kramers-Kronig transform, relying on our steady-state SE data of Fig. \ref{fig:Conductivity}(a) as a starting point and performing a Lorentz analysis of the $\Delta$R/R maps at different temperatures \cite{baldini2017clocking, baldini2017strongly}. As a result, the determination of the real part $\Delta\sigma_1$ gives access to the temporal evolution of the SW in the visible range.

\begin{figure}[tb]
	\includegraphics[width=\columnwidth]{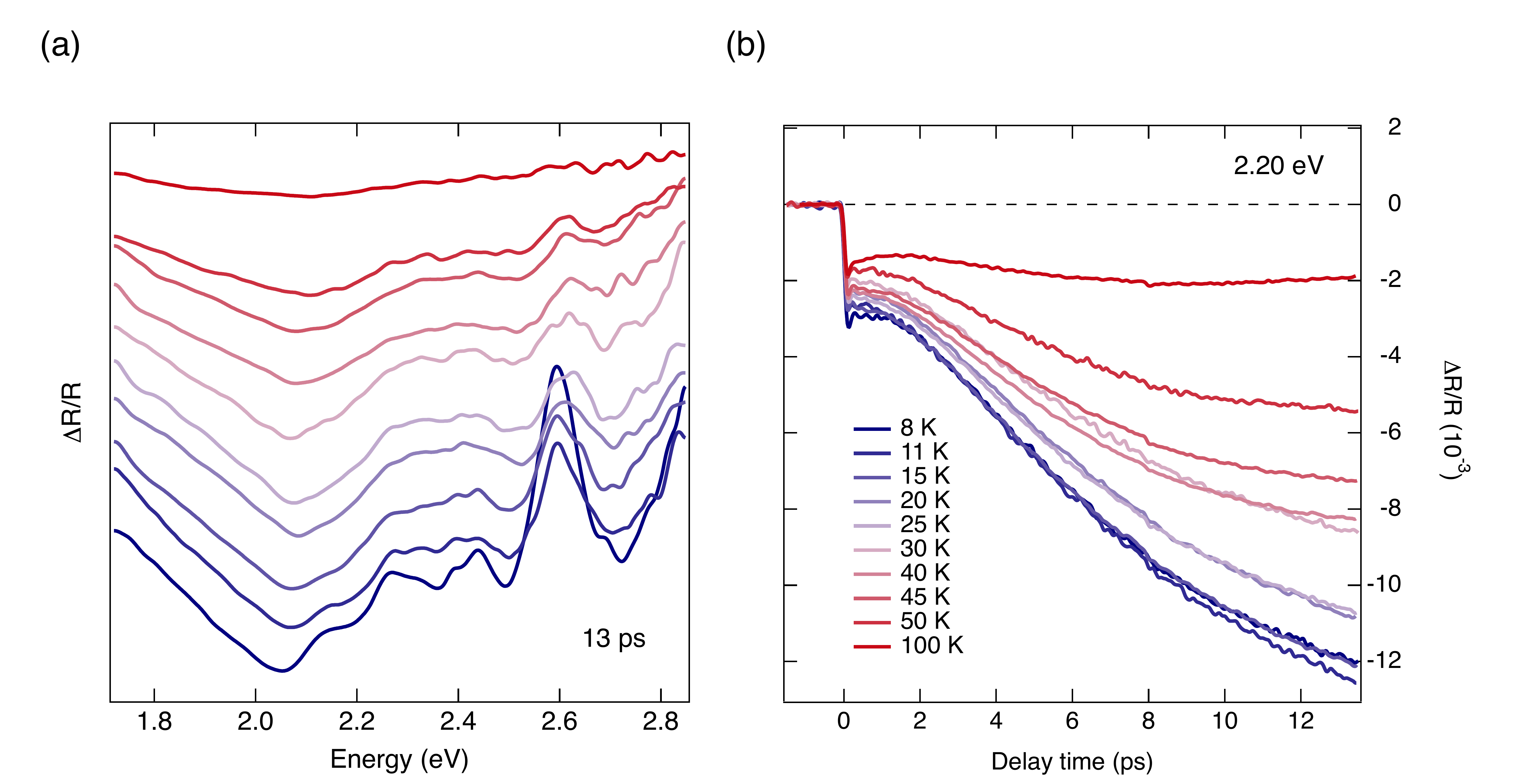}
	\caption{(a) Temperature dependence of the $\Delta$R/R  spectrum for a time delay of 13 ps. (b) Temperature dependence of the $\Delta$R/R temporal traces for a probe photon energy of 2.20 eV. The temperatures are indicated in the labels and all curves have been cut from the maps of Fig. S2.}
	\label{fig:Tdep_SpectrumTemporalTraces}
\end{figure}

To this aim, we perform a complete temperature dependence of $\Delta$R/R. The full set of data is shown in Fig. S2. We observe that the fine structure of the low-energy band manifesting at 8 K is gradually lost as the temperature is increased. A strong variation of the intensity of the response also occurs, and the measured changes become smaller than the noise level above 100 K. The fine structure of the low-energy optical band of TbMnO$_3$ becomes more evident when the long time delay $\Delta$R/R spectra are directly compared at different temperatures. These spectra are shown in Fig. \ref{fig:Tdep_SpectrumTemporalTraces}(a) for a time delay of 13 ps and are vertically shifted for clarity. The temporal evolution of the system around the probe photon energy of 2.20 eV is displayed in \ref{fig:Tdep_SpectrumTemporalTraces}(b) at different temperatures. Upon entering the magnetic phase below $\mathrm{T_{N1}}$, the dynamics slow down as observed previously \cite{ref:handayani}. This increase in the $\tau_m$ time constant for decreasing temperature was assigned to the signature of a photoinduced magnon-assisted hopping along the \textit{c}-axis of the material, which leads to an increase in the magnon number density and in turn affects both the \textit{a}- and \textit{c}-axis optical response. In the following, we will show that our data support instead a scenario in which an ultrafast lattice reorganization following the formation of small polarons is the source of the bottleneck observed in the spin order melting of TbMnO$_3$.

\begin{figure}[tb]
\centering
\includegraphics[width=\columnwidth]{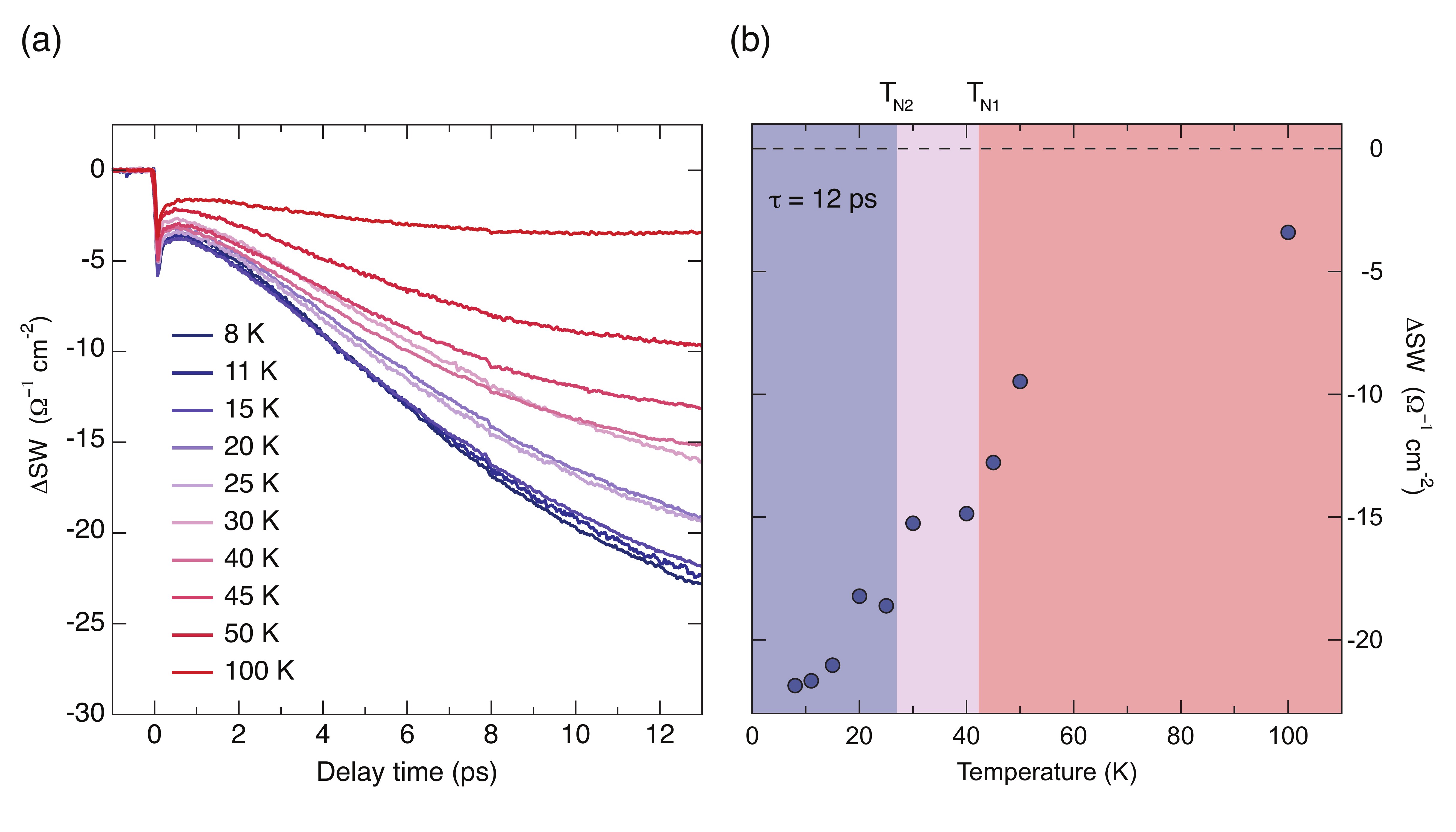}
\caption{(a) Comparison of the nonequilibrium $\mathrm{\Delta_{SW}}$ temporal dynamics at different temperatures, which are indicated in the label. At every time delay, the SW is calculated by computing the integral of the corresponding $\Delta\sigma_1$ map over the whole probed spectral range. The absorbed pump fluence is 4.4 mJ/cm$^2$. (b) Temperature evolution of the nonequilibrium SW integrated over the whole probe spectrum at 12 ps delay time. The blue shaded region highlights the temperature range where the material is in the multiferroic spin-cycloid phase, the violet region depicts the region where the SDW phase emerges and the red region represents the paramagnetic phase with short-range spin correlations. The respective temperature scales $\mathrm{T_{N2}}$ and $\mathrm{T_{N1}}$ are indicated on top.}
\label{fig:Temp_SW}
\end{figure}

From the temperature dependent $\Delta$R/R data, we calculate $\Delta\sigma_1$ as a function of probe photon energy and time delay. The color-coded maps at the different temperatures are shown in Fig. S3 and feature a prominent drop of $\Delta\sigma_1$ at large time delays. The determination of $\Delta\sigma_1$ at all temperatures allows us to follow the temporal evolution of the change in the partial SW ($\mathrm{\Delta_{SW}}$) over the probed range. The quantity $\mathrm{\Delta_{SW}}$ is defined as
\begin{equation}
\Delta_{SW} = \int_{\omega_1}^{\omega_2} \Delta\sigma_1 d\omega
\end{equation}
where $\omega_1$ = 1.72 eV and $\omega_2$ = 2.85 eV. A direct comparison among the $\mathrm{\Delta_{SW}}$ temporal dynamics at the different temperatures is established in Fig. \ref{fig:Temp_SW}(a). Here, our primary interest is to reveal whether the magnetic order dynamics gives rise to detectable temperature anomalies in $\mathrm{\Delta_{SW}}$ at long time delays. For this reason, we track the value of $\mathrm{\Delta_{SW}}$ at 12 ps for the different temperatures. The results are presented in Fig. \ref{fig:Temp_SW}(b). 
For clarity, we also indicate the $\mathrm{T_{N1}}$ and $\mathrm{T_{N2}}$ temperature scales by highlighting different temperature regions with distinct colors. Starting from high temperatures, we observe that $\mathrm{\Delta_{SW}}$ decreases its value when approaching the first magnetic phase transition at $\mathrm{T_{N1}}$ and decreases even further close to the second magnetic phase transition at $\mathrm{T_{N2}}$. Two kinks are observed in the proximity of $\mathrm{T_{N1}}$ and $\mathrm{T_{N2}}$. While the former had been already observed at the single-wavelength
$\Delta$R/R level \cite{ref:handayani}, the latter is a previously undetected feature that can be accessed when the whole spectral region of the intersite $d$-$d$ transitions is covered by the probe pulse. Although the presence of these kinks could be correlated with the magnetic order dynamics, this interpretation is not unambiguous. Irrespective of the origin of this SW loss, the carriers lost in the visible range are likely to be redistributed to high energies, in correspondence to the optical band at 5.15 eV \cite{ref:kovaleva_prl, ref:bastjan}.

\subsection{C. Coherent collective response}
\label{CoherentTMO}

The possible detection of a dynamical SW transfer associated with the loss of magnetic order in photoexcited TbMnO$_3$ provides new insights into the nonequilibrium response of the material. However, this observable represents an indirect effect of the order parameter melting on the optical properties of the system. As such, our measurements only suggest that a delayed transfer of thermal energy from the excited carriers to the spin system is at play in the material, consistent with previous experimental results \cite{ref:johnson, bothschafter_TMO}. Thus, the microscopic mechanism behind the magnetic order melting remains unclear. One scenario that has been envisioned behind the delayed energy transfer is related to relaxation  of  the  JT distortion, with the subsequent localization of the carriers in the form of small polarons \cite{ref:johnson}. The charge localization would hinder magnon-assisted hopping and therefore would require that the energy transfer to the spin system to be mediated by changes in the lattice structure. 

This represents a crucial aspect, since the polaronic behavior of the charge carriers is widely recognized as one of the peculiarities of the orthorhombic manganites \cite{ref:yamada_polaron_TMO, ref:shimomura, ref:daoud}. Indeed, doping carriers \cite{ref:mildner} or photoexciting the intersite \textit{d}-\textit{d} CT transitions in perovskite manganites is predicted to lead to the creation of the so-called anti-JT polarons \cite{ref:allen_TMO}. Similar effects are envisioned when the \textit{p}-\textit{d} CT transitions are photoexcited \cite{ref:mertelj_TMO}. Indeed, as the lattice shows a cooperative JT distortion, the presence of an extra charge (\textit{i.e.} electron or hole) on the Mn$^{3+}$ ions can produce a strong structural rearrangement to locally remove and relax the JT distortion. In this way, the system tries to minimize the energy cost generated by the existence of additional charge in a collectively JT-distorted crystal. This site acts as a defect that becomes strongly pinned, since the hopping to other Mn$^{3+}$ sites requires moving along the lattice distortion. In magnetically ordered phases, also the spin degree of freedom can be affected by the charge localization, as the polaron is expected to produce canting of the spins from their natural directions. Behaving similarly to a defect, the anti-JT polaron is a prototypical example of a small (Holstein) polaron and, as such, it involves the presence of a local deformation around the self-trapped carrier \cite{ref:holstein1, ref:holstein2}. Therefore, the local symmetry associated with the displacement is expected to retain a totally symmetric character ($\mathrm{A_{g}}$ symmetry).

To investigate the validity of this scenario, we search for the signatures of coherent optical phonon modes with totally symmetric character that are coupled to the photoexcited carriers and signal a rearrangement of the lattice structure consistent with the relaxation of the JT distortion. Here, we go beyond the results of Section III.A and monitor the \textit{a}-axis $\Delta$R/R at 8 K by decreasing the time step for the detection to $\sim$13 fs.

\begin{figure}[tb]
\centering
\includegraphics[width=\columnwidth]{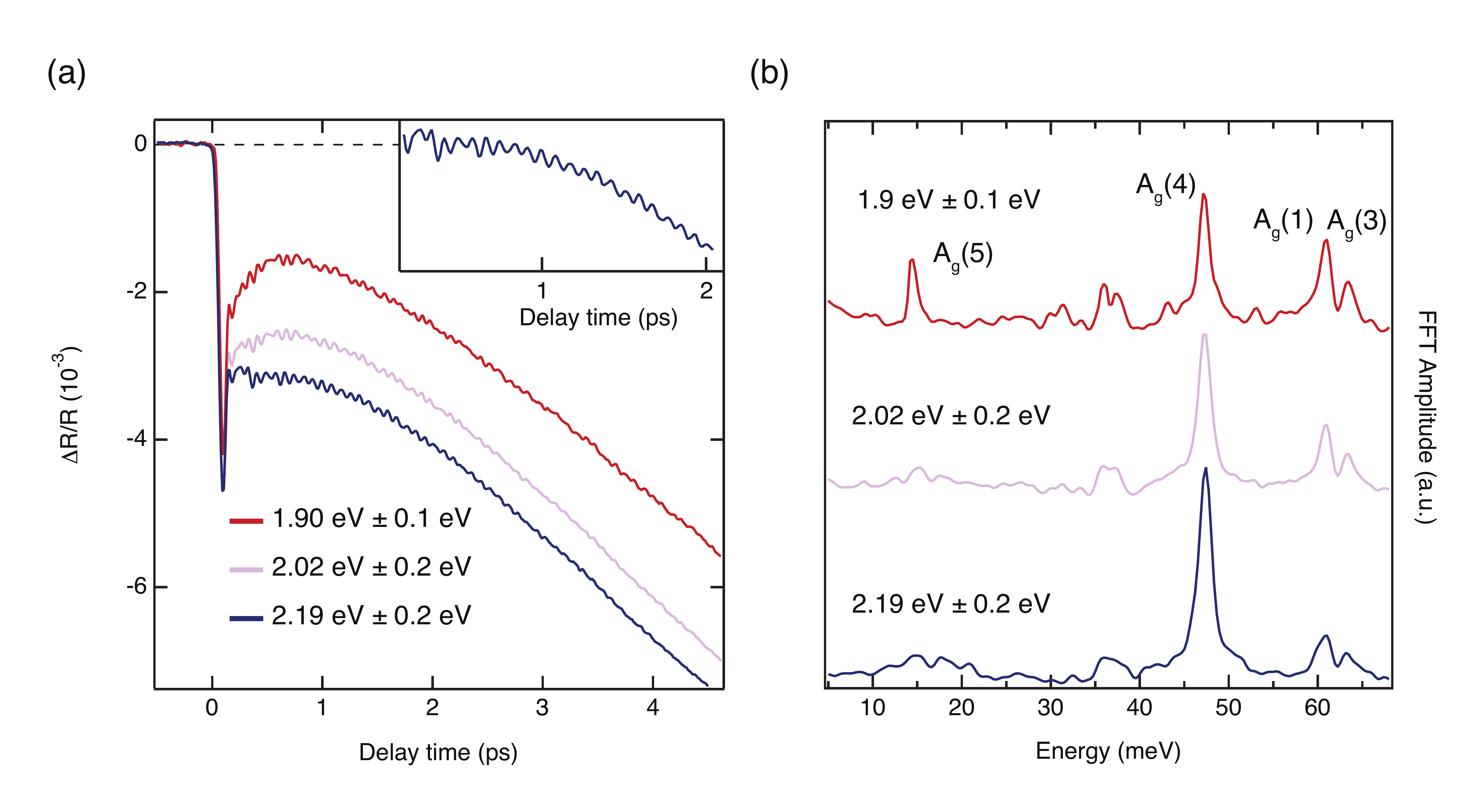}
\caption{(a) Temporal dynamics and (b) FT of the spectral response at 1.90 eV, 2.02 eV and 2.19 eV, averaged over the region indicated in the label. The pump polarization is set along the \textit{a}-axis. The absorbed pump fluence is 4.4 mJ/cm$^2$. Inset in (a) Details of the coherent oscillations on the trace at 2.19 eV between 0 and 2 ps. The assignment of the coherent modes is given in (b).}
\label{fig:Coherent_TMO}
\end{figure}

Figure \ref{fig:Coherent_TMO}(a) shows some representative temporal traces when the absorbed pump fluence for is 4.4 mJ/cm$^2$. The probe photon energies at which these traces have been selected are indicated in the labels. Remarkably, the initial ultrafast relaxation of the electronic response is now clearly resolved, displaying a sharp and well-defined negative peak. Simultaneously, a coherent beating among several modes emerges in the time-domain during the electronic relaxation and persists up to 4 ps (see the inset of Fig. \ref{fig:Coherent_TMO}(a)). By performing a Fourier transform (FT) analysis of the residuals from the fit (Fig. \ref{fig:Coherent_TMO}(b)), we identify the presence of four collective modes taking part to the coherent dynamics triggered by the intersite \textit{d}-\textit{d} excitation. Their energies correspond to 14.5 meV ($\mathrm{A_{g}(5)}$), 47.2 meV ($\mathrm{A_{g}(4)}$), 61.0 meV ($\mathrm{A_{g}(1)}$) and 63.4 meV ($\mathrm{A_{g}(3)}$) and are indicative of four of the seven Raman-active $\mathrm{A_g}$ modes of TbMnO$_3$ \cite{ref:martincarron, ref:iliev, ref:laverdiere, ref:kumar, ref:rovillain}. In particular, $\mathrm{A_{g}(5)}$ corresponds to a soft mode involving the displacement of the Tb$^{3+}$ ion, $\mathrm{A_{g}(4)}$ to the rotation of the MnO$_6$ octahedra, $\mathrm{A_{g}(1)}$ to the anti-stretching JT vibrations of the O atoms in the xz plane and $\mathrm{A_{g}(3)}$ to the bending of MnO$_6$ octahedra. On the other hand, the doublet structure centered around 35-37 meV in the FT may arise from the convolution of the peaks associated with the $\mathrm{A_{g}(2)}$ and $\mathrm{A_{g}(7)}$ modes, which are known to be strongly intermixed in TbMnO$_3$ \cite{ref:iliev}. Indeed, this feature is found to persist over the probed spectral range, despite becoming broader as the probe photon energy is tuned above 2.00 eV. For the purpose of our discussion, we neglect the presence of this feature and avoid any speculation in the absence of a clear spectroscopic observable. Based on the observed frequencies and mode assignments, we also remark that no signature of coherent excitations of magnetic origin is detected either in these temporal traces, or in those measured up to $\sim$14 ps. Thus, we confirm that only four coherent $\mathrm{A_{g}}$ phonon modes resonate in the proximity to the \textit{d}-\textit{d} HS intersite absorption band. The same modes are observed when the pump polarization is set along the $c$-axis of the material, thus promoting LS intersite \textit{d}-\textit{d} transitions (Fig. S4). The emergence of these coherent collective modes with a well-defined symmetry allows us to propose an explanation of the ultrafast magnetic order dynamics occurring in the spin cycloid phase of TbMnO$_3$.

In the past, a two-color pump-probe study with a time resolution of 10 fs revealed the presence of the $\mathrm{A_{g}(4)}$ and $\mathrm{A_{g}(1)}$ modes in the ultrafast response of LaMnO$_3$ \cite{ref:wall_manganite}. The $\mathrm{A_{g}(5)}$ and $\mathrm{A_{g}(3)}$ modes were not detected. In this experiment, the pump photon energy was tuned to be resonant with the intersite \textit{d}-\textit{d} transition, thus promoting the CT between two neighbouring Mn$^{3+}$ sites and creating locally Mn$^{2+}$-Mn$^{4+}$ sites. Moreover, a detailed temperature study of the $\Delta$R/R was performed in order to track the relevant parameters of the coherent modes, such as the oscillation amplitude and the damping rates. Surprisingly, both $\mathrm{A_{g}(4)}$ and $\mathrm{A_{g}(1)}$ modes were found to sharply increase their intensity below $\mathrm{T_N}$; they also showed a pronounced decrease of their damping rates when the temperature was reduced toward $\mathrm{T_N}$, followed by an enhanced decay below $\mathrm{T_N}$. Although this experiment did not provide any information on the ultrafast magnetic order dynamics occurring in the material, it was concluded that the generation mechanism of the two modes proceeds via the displacive excitation \cite{ref:zeiger}. The trigger mechanism of mode $\mathrm{A_{g}(1)}$ (anti-stretching JT mode) was explained by observing that the Mn$^{2+}$ and Mn$^{4+}$ ions are no longer JT active, leading to a relaxation of the JT distortion that launches the coherent mode. Importantly, as we will discuss later, the disruption of the regular, pure Mn$^{3+}$ arrangements has been addressed from the theory perspective \cite{ref:allen_TMO}, leading to the scenario of anti-JT polaron formation. The excitation of mode $\mathrm{A_{g}(4)}$ (out-of-phase rotation of the MnO$_6$ octahedra) was instead interpreted by invoking the Goodenough-Anderson-Kanamori rules \cite{ref:anderson_TMO, ref:goodenough1_TMO, ref:goodenough2_TMO, ref:kanamori}. In the excited state of the manganite, the presence of neighbouring Mn$^{3+}$-Mn$^{4+}$ sites gives rise to two empty $e_g$ orbitals. Hence, according to the Goodenough-Anderson-Kanamori rules, the exchange interaction J becomes negative. In a similar way, the presence of neighbouring Mn$^{2+}$-Mn$^{3+}$ sites leads to two half-filled $e_g$ levels, providing again a negative J. The change in sign of J results in the establishment of a force to reduce the Mn-O-Mn semicovalent bond length. This can be achieved by both the relaxation of the JT distortion and by reducing the bond angle to give a straighter bond, which in turn excite the coherent mode. In other words, the renormalization of J under nonequilibrium conditions triggers the coherent lattice motion via the displacive mechanism. 

As already observed above, modes $\mathrm{A_{g}(5)}$ and $\mathrm{A_{g}(3)}$ were not observed in the nonequilibrium dynamics of LaMnO$_3$, and thus they represent novel features detected by our measurement. The experimental parameters (time resolution, and pump/probe photon energy) used in Ref. \cite{ref:wall_manganite} were indeed suitable for revealing the presence, if any, of both modes. This suggests that these coherent modes are a peculiar feature of the ultrafast dynamics of TbMnO$_3$, which possesses a higher degree of distortion than LaMnO$_3$ and displays multiferroicity. Given the complex beating among the different modes, extracting the phase of each individual mode (to distinguish whether the temporal behavior is a sine or a cosine function) becomes a challenge. Hence, to explain the appearance of modes $\mathrm{A_{g}(5)}$ and $\mathrm{A_{g}(3)}$, we rely on additional considerations. Concerning mode $\mathrm{A_{g}(3)}$, we base our arguments on spontaneous Raman scattering data for the series of orthorhombic RMnO$_3$ manganites. It was found that the stretching $\mathrm{A_{g}(1)}$ and bending $\mathrm{A_{g}(3)}$ modes, while uncorrelated for R = La, display a pronounced mixing for R = Sm, Eu, Gd, Tb \cite{ref:iliev}. This effect is enhanced by the proximity between their phonon energies when the unit cell becomes highly distorted. Two phonon modes of same symmetries and close energies can be considered as coupled oscillators, whose frequencies are given by
\begin{equation}
\hbar\omega_{1,2} = \frac{\hbar\omega^{\prime} + \hbar\omega^{\prime\prime}}{2} \pm \sqrt{\frac{(\hbar\omega^{\prime} - \hbar\omega^{\prime\prime})^2}{4} + \frac{V^2}{4}}
\end{equation}
where $\hbar\omega^{\prime}$ and $\hbar\omega^{\prime\prime}$ are the mode energies without coupling and V is the coupling constant. From spontaneous Raman scattering, V can be estimated $\sim$2.5 meV. This leads us to conclude that, in our nonequilibrium experiment, the coherent excitation of mode $\mathrm{A_{g}(3)}$ is strictly connected to the coherent excitation of mode $\mathrm{A_{g}(1)}$, as the two modes are mutually dependent.

The generation mechanism of the coherent $\mathrm{A_{g}(5)}$ mode is instead more subtle. As discussed above, this mode corresponds to the vibration of the Tb$^{3+}$ ion along the \textit{c}-axis and was identified as a partially softened phonon mode associated with the ferroelectric transition \cite{ref:rovillain}. In the past, ferroelectric phase transitions have been widely investigated via pump-probe spectroscopy to track the dynamics of coherent soft modes in order-disorder type perovskites (such as KNbO$_3$ and SrTiO$_3$) \cite{ref:dougherty_1992, ref:dougherty_1994, ref:kohmoto} and in displacive type ferroelectrics (such as GeTe) \cite{ref:hase}. While in the first class of materials the generation mechanism of the soft modes has been associated with Impulsive Stimulated Raman Scattering (ISRS) \cite{ref:merlin, ref:testevens}, in the second class a displacive excitation has been proposed \cite{ref:zeiger}. In our experiment, the mode is found to strongly resonate only in the low-energy wing of the visible spectrum (around 1.90 eV). In this region, the absorptive part of the optical conductivity is weak and featureless, suggesting that the displacive character of the exctiation is weak. Consistent with this hypothesis, we observe that the 1.90 eV region in which the $\mathrm{A_{g}(5)}$ mode resonates corresponds to the pronounced dip found in the dispersive part of the optical conductivity (Fig. S1(b)). This leads us to conclude that the ISRS scenario could explain the excitation mechanism of this partially softened mode in TbMnO$_3$. 

The observation of coherent optical phonons in the nonequilibrium dynamics of TbMnO$_3$ opens intriguing perspectives in the evaluation of the electron-phonon coupling for all these modes. Following this consideration, a specific and quantitative estimate of the coupling between the electrons and the coherent Raman-active $\mathrm{A_g}$ optical phonons via \textit{ab initio} calculations will provide insightful information, albeit projected at the $\Gamma$ point. We will address this aspect in a future publication.

\begin{figure}[tb]
\centering
\includegraphics[width=\columnwidth]{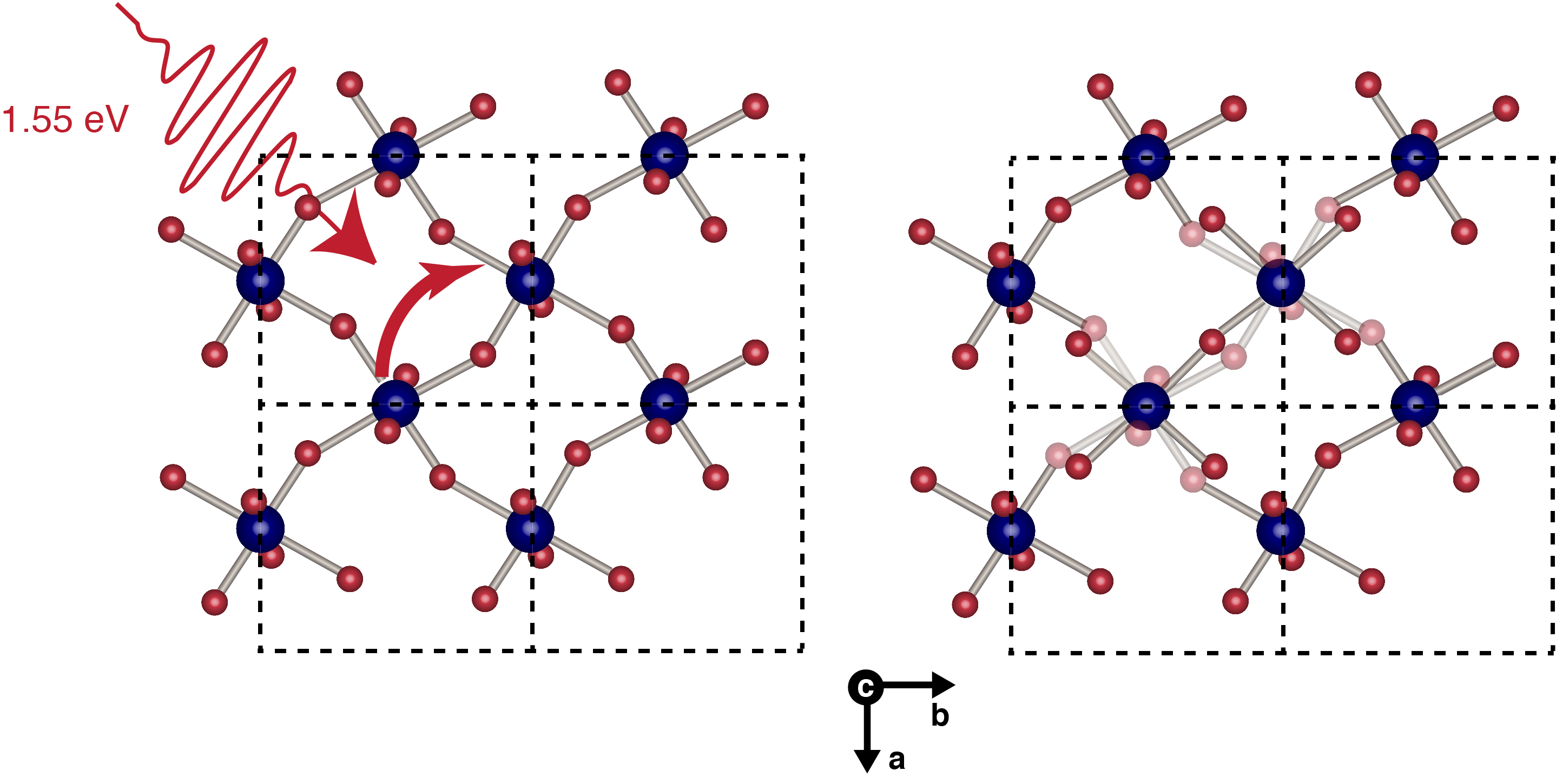}
\caption{Schematic illustration of the lattice displacement associated with the creation of an anti-JT polaron upon photoexcitation of the system via an intersite \textit{d}-\textit{d} CT transition.}
\label{fig:AJT}
\end{figure}

\section{IV. Conclusions}
\label{Conclusion}

In this work we have applied ultrafast optical spectroscopy over a broad wavelength range to study the sequence of events leading to spin-order melting in laser-excited multiferroic TbMnO$_3$. The interaction between the pump pulse at 1.55 eV and the system leads to the excitation of an intersite \textit{d}-\textit{d} CT transition, corresponding to an optical excitation across the fundamental Mott-Hubbard gap. In other words, the excitation locally promotes the creation of Mn$^{2+}$-Mn$^{4+}$ sites, leading to the disruption of the regular Mn$^{3+}$ arrangement and to the relaxation of the JT distortion. In this scenario, anti-JT small polarons are expected to form, producing a local distortion in the MnO$_6$ octahedra \cite{ref:allen_TMO}. A pictorial representation of the photoexcitation process and the formation of anti-JT polarons is given in Fig. \ref{fig:AJT}. The change in the lattice structure is reflected in the emergence of the coherent $\mathrm{A_g(1)}$ anti-stretching JT mode and of the $\mathrm{A_g(3)}$ bending mode, to which the $\mathrm{A_g(1)}$ is strongly mixed. Simultaneously, the photoexcited charge density couples via the exchange interaction to the $\mathrm{A_g(4)}$ mode, corresponding to the out-of-phase rotation of the MnO$_6$ octahedra. Also this structural mode may be involved in the distortion associated with the anti-JT polaron formation. We propose that these coherent modes represent the signatures of the creation of anti-JT polarons in TbMnO$_3$. The formation of such a relatively long-lived self-trapped charge hinders magnon-assisted hopping and thus requires  the  energy  transfer  to  the  spin  system to be mediated  by  the rearrangement  of  the  lattice  structure. Consistent with this idea, the signal associated with the melting of the long-range magnetic order rises within several ps and, at our absorbed pump fluence, the complete melting of the magnetic order is expected to take place within 22 ps \cite{ref:johnson}. We expect polaronic effects to manifest also when the initial photoexcitation couples to the \textit{p}-\textit{d} CT transition, as the $d$-orbitals will be again influenced by the presence of an extra charge. Similar experiments exploring a pump photon energy at 3.10 eV confirm the insensitivity of the magnetic order response to the details of the initial photoexcitation \cite{ref:handayani, bothschafter_TMO}. 

The scenario proposed above reconciles the results reported in different experiments, suggesting the crucial involvement of the lattice behind the spin-order melting. More importantly, as the coherent optical phonons were previously detected in LaMnO$_3$, it is reasonable to believe that the same phenomenology is effectively at play in all undoped orthorhombic manganites. Indeed, the pronounced coherences appear in both LaMnO$_3$ and TbMnO$_3$, which represent the least and the most distorted orthorhombic manganites of the RMnO$_3$ family, respectively. A different microscopic mechanism may be at play in the hexagonal manganites of the RMnO$_3$ family, as the hexagonal crystal field splitting lifts the \textit{d}-orbital degeneracy in a different fashion \cite{ref:rai_TMO, ref:souchkov}. In this case, a pump pulse at 1.55 eV promotes an \textit{on-site} \textit{d}-\textit{d} transition of the Mn electron, which leaves the total charge unchanged and does not modify the polaronic potential \cite{ref:bowlan}. 

In addition, two previously undetected coherent phonon modes are found to participate to the ultrafast evolution of the system after the photoexcitation. In particular, the $\mathrm{A_g(5)}$ mode corresponds to the partially softened phonon associated with the ferroelectric phase transition. We speculate that this mode is generated via the ISRS mechanism, similarly to the behavior observed in the order-disorder perovskite ferroelectrics \cite{ref:dougherty_1992, ref:dougherty_1994, ref:kohmoto}. We suggest that temperature dependent studies of this partially softened mode may unveil new intriguing aspects on the origin of the magnetoelectric coupling in this multiferroic material and shed light on the fate of the ferroelectric order parameter following the photoexcitation. One interesting open question is whether the ferroelectric and the magnetic order parameters decouple simultaneously after the interaction with the pump pulse and follow separate temporal evolution in the system. In this regard, our measurements set the basis for revealing the dynamics of the ferroelectric polarization in the spin-cycloid magnetic phase with more sensitive ultrafast methods, such as time-resolved second harmonic generation spectroscopy \cite{Mankowsky} and microscopy \cite{matsubara2015magnetoelectric}.

\begin{acknowledgments}
We acknowledge A. Mann and S. Borroni for the technical support. Work at LUMES was supported by the Swiss National Science Foundation through the NCCR MUST. Crystal growth work at the Institute for Quantum Matter (IQM) was supported by the U.S. Department of Energy, Office of Basic Energy Sciences, Division of Materials Sciences and Engineering through Grant No. DE-FG02-08ER46544.
\end{acknowledgments}

\noindent* \email[{ebaldini@mit.edu}

\clearpage
\newpage

\setcounter{section}{0}
\setcounter{figure}{0}
\renewcommand{\thesection}{S\arabic{section}}  
\renewcommand{\thetable}{S\arabic{table}}  
\renewcommand{\thefigure}{S\arabic{figure}} 
\renewcommand\Im{\operatorname{\mathfrak{Im}}}

\section{S1. Experimental methods}

\subsection{A. Sample preparation}

A high quality stoichiometric TbMnO$_3$ single crystal was produced by the optical floating zone technique at the zoning rate of 0.5 mm/h with rotation rate of 15 rpm for the growing crystal and 0 rpm for the feed rod under static argon. The crystal was oriented using Laue backscattering and cut to expose the (010) face with an approximately 3$^\circ$ miscut. The surface of the sample was polished and afterwards the sample was annealed in air in 650$^\circ$C for 110 hours. The dimensions of the crystal are approximately 2 mm $\times$ 2 mm $\times$ 3 mm along the \textit{a}, \textit{b} and \textit{c} axes, respectively. The Pbnm orthorhombic convention is used to describe the crystal axes.

\subsection{B. Spectroscopic ellipsometry}

We used spectroscopic ellipsometry to measure the complex dielectric function of the sample, covering the spectral range from 0.50 eV to 6.00 eV. The measurements were performed using a Woollam VASE ellipsometer. The TbMnO$_3$ single crystal was mounted in a helium flow cryostat, allowing measurements from room temperature down to 10 K. The measurements were performed at $<$10$^{-8}$ mbar to prevent measurable ice-condensation onto the sample. Anisotropy corrections were performed using standard numerical procedures \cite{ref:aspnes}.

\subsection{C. Ultrafast broadband optical spectroscopy}

Femtosecond broadband transient reflectivity experiments were performed using a set-up described in details in Ref. \cite{baldini2016versatile}. Briefly, a Ti:Sapphire oscillator, pumped by a continuous-wave Nd:YVO$_4$ laser, emitted sub-50 fs pulses at 1.55 eV with a repetition rate of 80 MHz. The output of the oscillator seeded a cryo-cooled Ti:Sapphire amplifier, which was pumped by a Q-switched Nd:YAG laser. This laser system provided $\sim$ 45 fs pulses at 1.55 eV with a repetition rate of 3 kHz. One third of the output, representing the probe beam, was sent to a motorized delay line to set a controlled delay between pump and probe. The 1.55 eV beam was focused on a 3 mm-thick CaF$_2$ cell using a combination of a lens with short focal distance and an iris to limit the numerical aperture of the incoming beam. The generated continuum covered the 1.72 - 2.85 eV spectral range. The probe was subsequently collimated and focused onto the sample through a pair of parabolic mirrors under an angle of 15$^\circ$. The remaining two thirds of the amplifier output, representing the pump beam, were directed towards the sample under normal incidence. Along the pump path, a chopper with a 60 slot plate was inserted, operating at 1.5 kHz and phase-locked to the laser system. Both pump and probe were focused onto the sample with spatial dimensions of \mbox{120 $\mathrm{\mu m}$ $ \times$ 87 $\mathrm{\mu m}$} for the pump and 23 $\mathrm{\mu m}$ $\times$ 23 $\mathrm{\mu m}$ for the probe. The sample was mounted inside a closed cycle cryostat, which provided a temperature-controlled environment in the range 10 - 340 K. The reflected probe was dispersed by a fiber-coupled 0.3 m spectrograph and detected on a shot-to-shot basis with a complementary metal-oxide-semiconductor linear array.

\section{S2. Computational details}

To assign the features appearing in the optical spectra, we performed \textit{ab initio} Density-Functional Theory (DFT) calculations of TbMnO$_3$. The electronic structure and optical properties of TbMnO$_3$ were carried out via Wien2k code. The exchange-correlation potential were treated using GGA + U, where U was set at 3 eV and 6 eV for Mn 3\textit{d} and Tb 4\textit{f} orbitals, respectively. The muffin-tin radii Rmt were 2.2, 1.95, 1.5 a.u. for Tb, Mn, and O atoms, respectively. The maximum angular momentum of the radial wavefunctions was set to 10, and RmtKmax was fixed at 7.0 to determine the basis size.

\section{S3. Additional results}



\subsection{A. Fit of the $\Delta$R/R data}

To get more quantitative information on the timescales governing the ultrafast dynamics, we perform a fit of some representative $\Delta$R/R traces at 8 K. For this purpose, we select the temporal traces around a probe photon energy of 1.80 eV, as in this region the short-lived component becomes more apparent. These traces are shown in Figs. 6(a,b) of the main text as violet curves along the \textit{a}- and \textit{c}-axis, respectively. Among the several models that can be implemented for capturing the dynamics, the simplest one consists of three exponential functions convolved with a Gaussian accounting for the temporal shape of the pump pulse. While the shortest lived component appears immediately after the pump photoexcitation, the other two components display a slow rise. The function that is used for fitting the data is
\begin{widetext}
	\begin{equation}
	f(t) = u(t) \Bigg[f_{1}(t) + f_{2}(t) + f_{m}(t)\Bigg] = u(t) \Bigg[e^{\frac{-t^2}{\tau_{R_1}}} \ast A_{1} e^{-\frac{t-t_{D_1}}{\tau_{1}}} + \Big( 1-e^{\frac{-t^2}{\tau_{R_2}}}\Big) \ast \Big[
	A_{2} e^{-\frac{t-t_{D_2}}{\tau_{2}}} + A_{m} e^{-\frac{t-t_{D_2}}{\tau_{m}}}\Big]\Bigg]
	\label{TMO1}
	\end{equation}
\end{widetext}
\noindent where u(t) is a step function that has u = 0 for t $<$ 0 and u = 1 for t $\geq$ 0, $\mathrm{A_{1}}$, $\mathrm{A_{2}}$ and $\mathrm{A_{m}}$ are the amplitudes of the three exponential functions; $\tau_{R_1}$ and $\tau_{R_2}$ are the rise times of the exponential functions; $\tau_{1}$, $\tau_{2}$ and $\tau_{m}$ are the relaxation constants of the three exponentials; $t_{D_1}$ and $t_{D_2}$ are delay parameters with respect to the zero time. While all three exponential functions are necessary for fitting the \textit{a}-axis response, only two of them are sufficient to reproduce the \textit{c}-axis dynamics. The results of the fit are shown in Figs. 6(a,b) as black lines superposed on the original data. The timescales retrieved along the \textit{a}-axis are $\tau_{1}$ = 120 $\pm$ 10 fs, $\tau_{2}$ = 3 $\pm$ 0.7 ps and $\tau_{m}$ = 9 $\pm$ 1.5 ps, while those along the \textit{c}-axis are $\tau_{1}$ = 2 $\pm$ 0.5 ps and $\tau_{m}$ = 9 $\pm$ 0.5 ps. A larger uncertainty is expected on $\tau_{m}$ due to the limited temporal window of $\sim$12 ps probed in our experiment. 

\begin{figure*}[h]
	\centering
	\includegraphics[width=1.3\columnwidth]{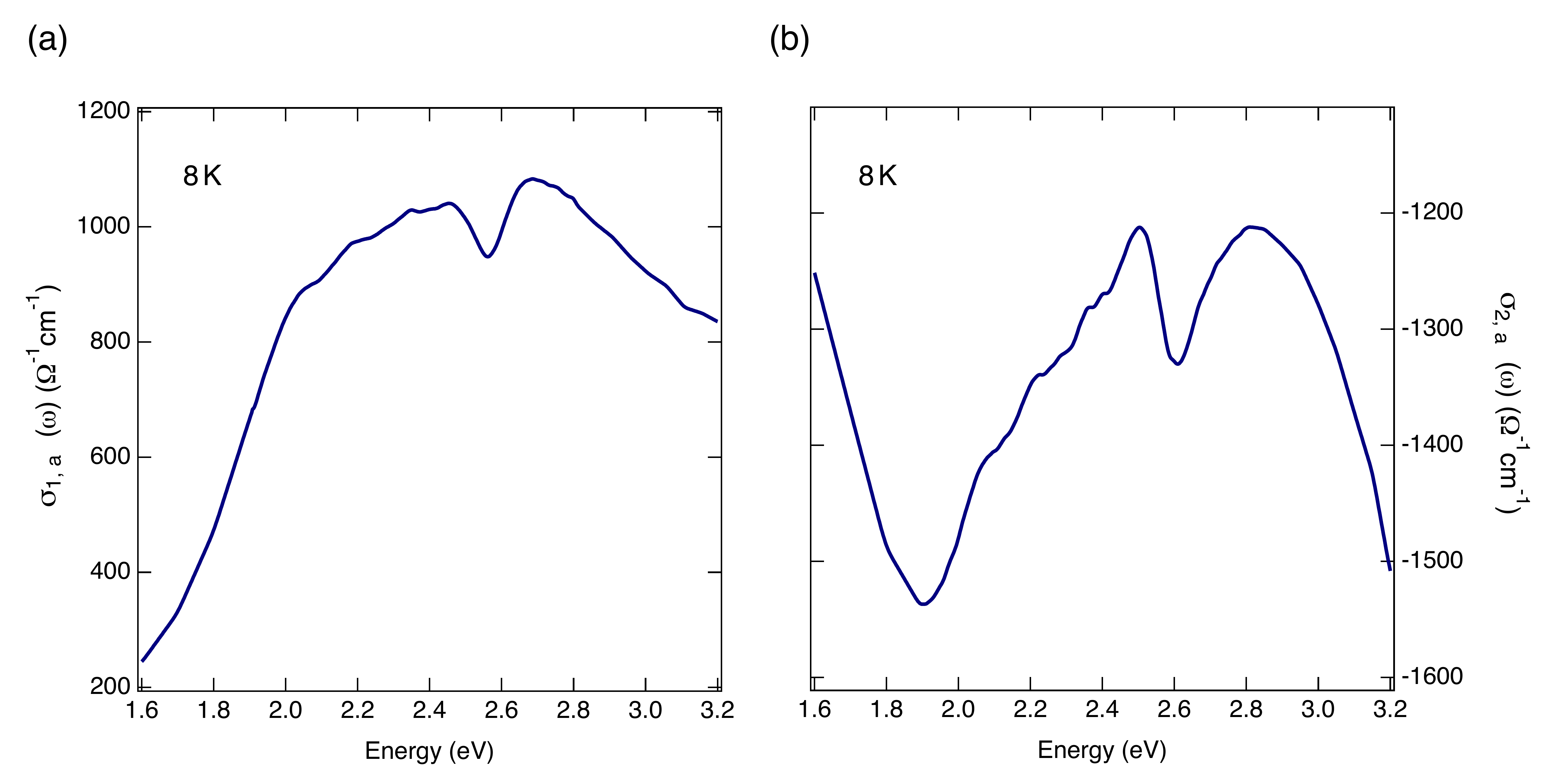}
	\caption{(a) Real $\sigma_{1a}$($\omega$) and (b) imaginary $\sigma_{2a}$($\omega$) parts of the \textit{a}-axis complex optical conductivity of TbMnO$_3$, measured at 8 K via SE.}
	\label{fig:HighResolution_SE}
\end{figure*}

\begin{figure*}[h]
	\centering
	\includegraphics[width=2\columnwidth]{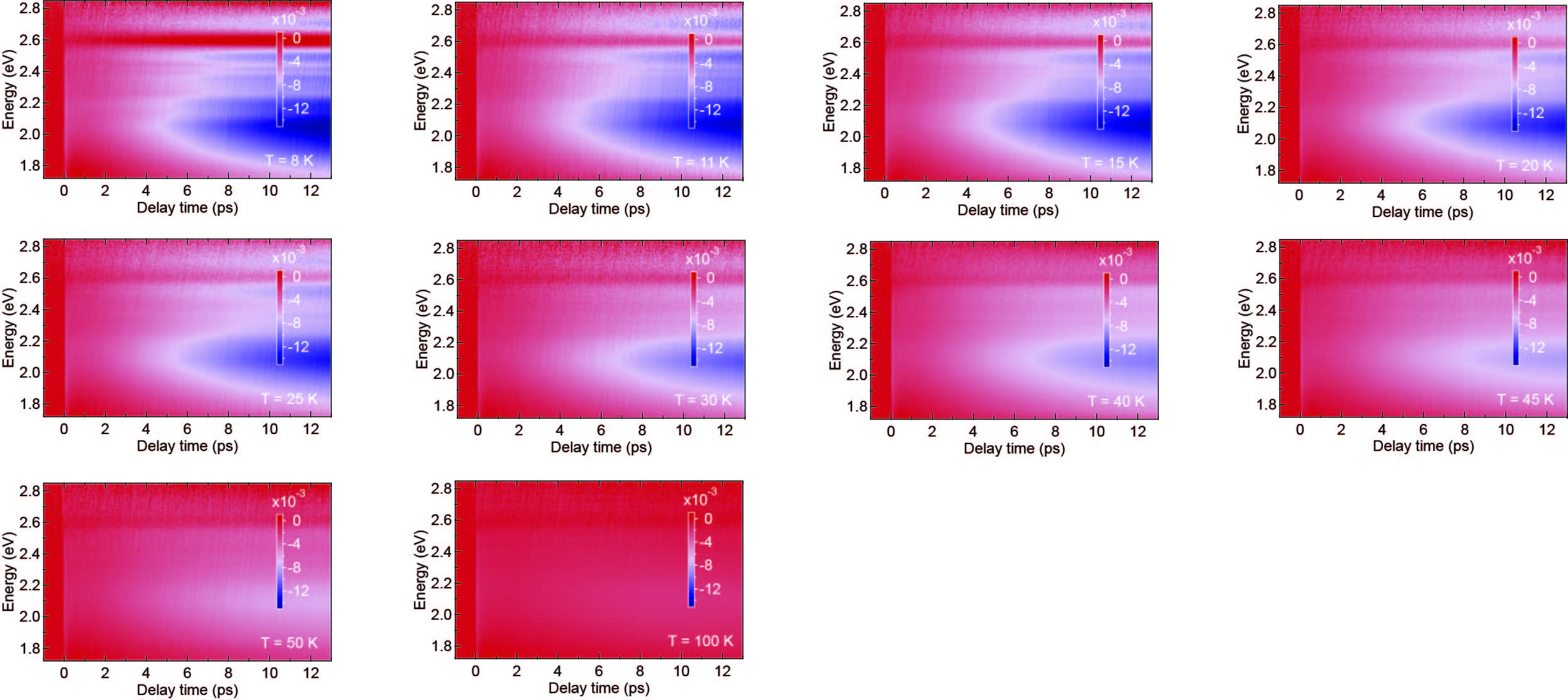}
	\caption{Temperature dependence of $\Delta$R/R as a function of probe photon energy and time delay between pump and probe. The temperatures are indicated in the labels and the absorbed pump fluence is 4.4 mJ/cm$^2$.}
	\label{fig:TransientReflectivity_TMO}
\end{figure*}

\subsection{B. Temperature dependence}

\noindent To monitor how the spectral signature of the \textit{a}-axis HS band varies in the nonequilibrium experiment across the two magnetic phase transitions, we perform a complete temperature dependence of $\Delta$R/R. The color-coded maps of $\Delta$R/R are shown in Fig. \ref{fig:TransientReflectivity_TMO}, in which the different temperature values are also indicated. We observe that the fine structure of the low-energy band manifesting at 8 K is gradually lost as the temperature is increased. A strong variation of the intensity of the response also occurs and the signal declines to the sensitivity range of our setup around the temperature of 100 K. The fine structure of the low-energy optical band of TbMnO$_3$ becomes more evident when the long time delay $\Delta$R/R spectra are directly compared at different temperatures. 

\begin{figure*}[tb]
	\centering
	\includegraphics[width=2\columnwidth]{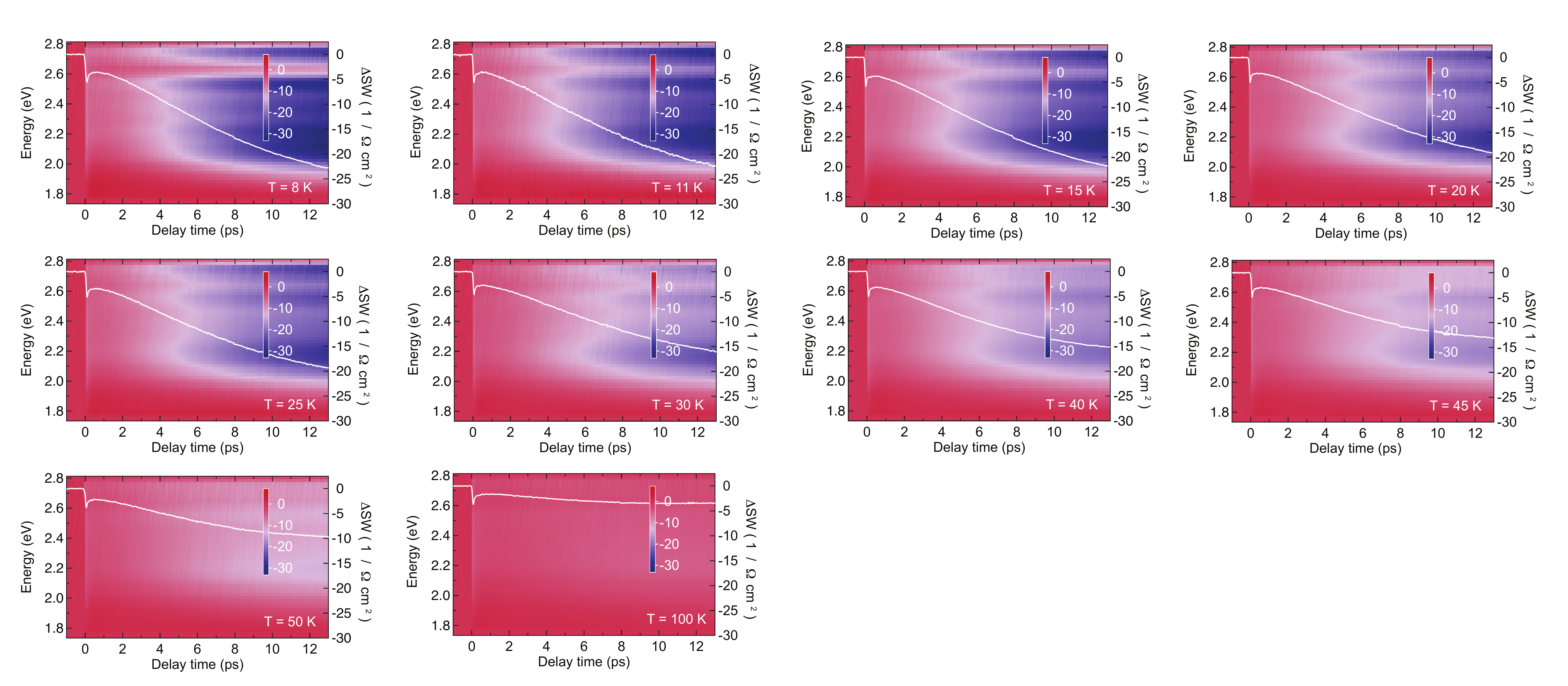}
	\caption{Temperature dependence of the transient optical conductivity $\Delta\sigma_1$ as a function of probe photon energy and time delay between pump and probe. Every map also shows the temporal evolution of the nonequilibrium SW ($\mathrm{\Delta_{SW}}$), which is calculated by computing the integral of the corresponding $\Delta\sigma_1$ map over the whole probed range. The temperatures are indicated in the labels and the absorbed pump fluence is 4.4 mJ/cm$^2$.}
	\label{fig:TransientConductivity_TMO}
\end{figure*}

A useful quantity that has to be extracted from the nonequilibrium experiment is the transient complex optical conductivity $\Delta\sigma$ = $\Delta\sigma_1$ + i $\Delta\sigma_2$. This can be calculated without the need of a Kramers-Kronig transform by relying on our steady-state SE data of Fig. 2(a) as a starting point and performing a Lorentz analysis of the $\Delta$R/R maps at the different temperatures. As a consequence, the determination of the real part $\Delta\sigma_1$ gives access to the temporal evolution of the spectral weight (SW) in the visible range. In Fig. \ref{fig:TransientConductivity_TMO} we show the calculated $\Delta\sigma_1$ at all temperatures, as a function of probe photon energy and time delay. At low temperatures, a prominent drop dominates the whole spectral range especially at large time delays. As the temperature increases, the absolute strength of the response becomes smaller and declines to the order of our sensitivity close to 100 K. The determination of $\Delta\sigma_1$ at all temperatures allows following the temporal evolution of the change in SW ($\Delta$SW) over the probed range. The dynamics of $\Delta$SW are overlapped to all color-coded maps of $\Delta\sigma_1$ in Fig. \ref{fig:TransientConductivity_TMO}. 

\begin{figure*}[h]
	\centering
	\includegraphics[width=1.3\columnwidth]{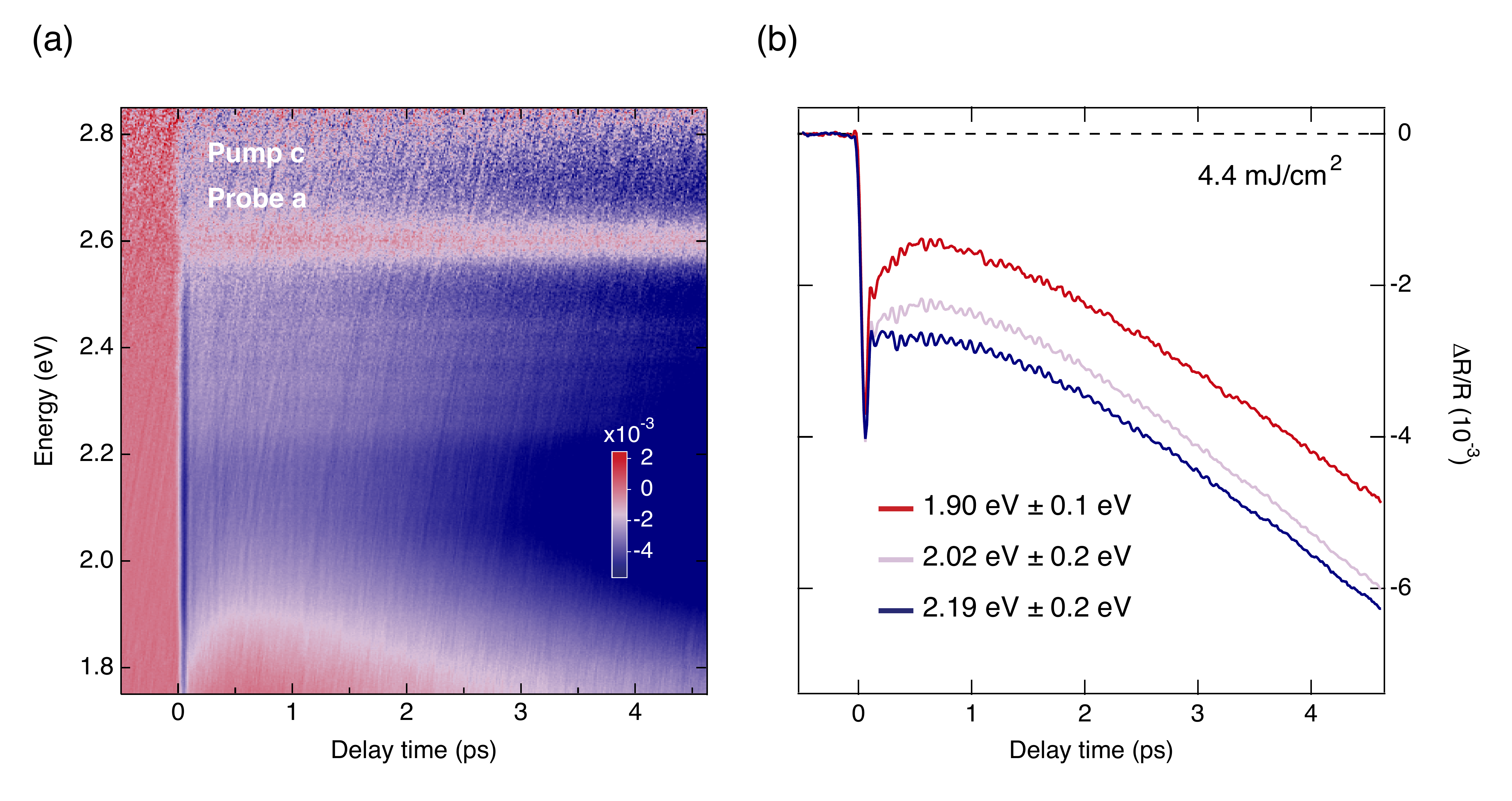}
	\caption{(a) Color-coded maps of $\Delta$R/R at 8 K with \textit{c}-axis pump polarization and \textit{a}-axis probe polarization. The pump photon energy is 1.55 eV and the absorbed pump fluence is 4.4 mJ/cm$^2$. (b) Temporal dynamics of the spectral response at 1.90 eV, 2.02 eV and 2.19 eV, averaged over the region indicated in the label.}
	\label{fig:Coherent_ca_TMO}
\end{figure*}

\subsection{C. Pump polarization along the c-axis}

To test whether the coherent modes modulating the \textit{a}-axis reflectivity can be excited also under other pump polarization conditions, we perform a separate experiment at 8 K where the pump beam is polarized along the \textit{c}-axis. In these conditions, the pump photon energy at 1.55 eV can promote intersite \textit{d}-\textit{d} transitions along the \textit{c}-axis, as it is resonant with the tail of the \textit{c}-axis low-spin \textit{d}-\textit{d} absorption feature shown in Fig. 2(c). The color-coded map of the \textit{a}-axis $\Delta$R/R is displayed in Fig. \ref{fig:Coherent_ca_TMO}(a) as a function of probe photon energy and time delay between pump and probe. Despite the weak \textit{c}-axis absorption, we retrieve a sizable $\Delta$R/R signal, retaining the same spectral shape of the response in Figs. 5(a,b) and 10(a). Some representative temporal traces selected from the map are shown in Fig. \ref{fig:Coherent_ca_TMO}(b) and demonstrate the emergence and persistence of the coherent response also for a pump polarization along the \textit{c}-axis. The Fourier transform analysis of the residuals from a fit of the incoherent response (not shown) confirms the presence of all previously listed phonon modes.

\newpage

\providecommand{\noopsort}[1]{}\providecommand{\singleletter}[1]{#1}%

\end{document}